\documentclass[11pt,a4paper]{article}
\usepackage{epsfig,graphicx,amsfonts,amsmath}

\usepackage{cite}
\RequirePackage{mathrsfs}

\newlength{\dinwidth}
\newlength{\dinmargin}
\def\eq#1{{Eq.~(\ref{#1})}}
\newcommand{\Le}{\left(}
\newcommand{\Ra}{\right)}

\newcommand{\beq}{\begin{equation}}
\newcommand{\eeq}{\end{equation}}
\newcommand{\beqar}{\begin{eqnarray}}
\newcommand{\eeqar}{\end{eqnarray}}
\newcommand{\D}{\partial}

\newcommand{\ep}{\varepsilon}

\date{}
\begin{document}

\title {{~}\\
{\Large \bf  One loop light-cone QCD, effective action for reggeized gluons and QCD RFT calculus}\\}
%
\author{ 
{~}\\
{\large 
S.~Bondarenko$^{(1) }$,
L.~Lipatov$^{(2)}$,
S.~Pozdnyakov$^{(1) }$,
A.~Prygarin$^{(1) }$
}\\[7mm]
{\it\normalsize  $^{(1) }$ Physics Department, Ariel University, Ariel 40700, Israel}\\
{\it\normalsize  $^{(2) }$ St.Petersburg State University, St. Petersburg 199034}\\
{\it\normalsize and Petersburg Nuclear Physics Institute, Gatchina 188300, Russia}\\
}

\maketitle
\thispagestyle{empty}

\begin{abstract}

 The effective action for reggeized gluons is based on the gluodynamic Yang-Mills Lagrangian  with  external current for longitudinal gluons added, 
see \cite{LipatovEff}. 
On the base of classical solutions,  obtained in \cite{Our1}, the
one-loop corrections to this effective action in light-cone gauge are calculated. The
RFT calculus for reggeized gluons similarly to the RFT introduced in \cite{Gribov}
is proposed and discussed. The correctness of the results is verified by calculation
of the propagator of $A_{+}$ and $A_{-}$ reggeized gluons  fields and
application of the obtained results is discussed as well.

\end{abstract}

\section{Introduction}

$\,\,\,\,\,\,$ The action for interaction of reggeized gluons  was introduced in the series of papers \cite{LipatovEff} and  describes multi-Regge processes 
at high-energies, see \cite{EffAct}. There are the following important applications of this action: it can be used for the calculation of 
production amplitudes in  different scattering processes and calculation of sub-leading, unitarizing  corrections to the  
amplitudes and production vertices, see  \cite{LipatovEff, EffAct, BKP, GLR, BK}. The last task can be considered as a construction of the RFT (Regge Field Theory) based on the interaction of the fields of reggeized gluons, where different vertices of
the interactions are introduced and calculated. 
The phenomenological RFT based on the Pomeron degrees of freedom was introduced in \cite{Gribov}. From this point of view we consider the effective action for
reggeized gluons as RFT calculus based on the degrees of freedom expressed through colored reggeon fields (reggeized gluons).

 The construction of RFT based on QCD Lagrangian requires the knowledge of solutions of classical equations of motion in terms of reggeon fields $A_{+}$ and $A_{-}$.
Inserting these solutions again in the Lagrangian we can develop field theory fully in terms of reggeon fields,
with loops corrections to the action determined in terms of these fields as well. Subsequent expansion of the action in terms of the reggeon fields will produce all possible
vertices of interactions of the fields, with precision determined by the precision of calculations in the framework of the QFT. 
From the QFT point of view, therefore, the problem of interest is the calculation of the one-loop effective action for gluon QCD Lagrangian with added external current
by use of the the non-trivial classical solutions expressed in terms of new degrees of freedom, see \cite{Our1}. 
These calculations
of the one-loop corrections to the effective action we perform in light-cone gauge using classical 
solutions from \cite{Our1}. The correctness of the obtained results can be checked by calculations of functions which are well-known in small-x
BFKL approach, \cite{BFKL}. The basic function there is the gluon Regge trajectory, which determines the form of the propagator of reggeized gluon fields
$A_{+}$ and $A_{-}$. This propagator in the proposed framework can be considered as operator inverse to the effective vertex of interaction of reggeon fields,
this check is performed in the paper.
There are also
other possibilities to verify the self-consistency of the approach. For example, it can be calculation of the BFKL kernel, which is 
an effective vertex of interactions of four reggeons, or  calculation of the triple Pomeron vertex, see \cite{TripleV}, which is interaction vertex of six reggeon fields. These calculations will be considered further in separate publications.

  Thus, below, we calculate one-loop effective action for reggeized gluons and calculate propagator for $A_{+}$ and $A_{-}$ reggeon fields. Respectively, in the Section 2, 
we remind main results obtained for the classical solution of the effective action for reggeized gluons. In Section 3 we consider the expansion of the Lagrangin in terms of fluctuations around these classical solutions, whereas in Section 4 we calculate the one-loop correction	to the classical action obtaining effective one loop action in terms of reggeon fields. 
In Section 5 we discuss RFT calculus based on the effective action and in Section 6 we verify the correctness of the result calculating the propagator of the reggeized gluons fields.
Section 7 is the conclusion of the paper, there are also Appendixes A, B, C where the main calculations related to the result are presented.

\section{Effective action for reggeized gluons and classical equations of motion}

$\,\,\,\,\,\,$ The effective action, see \cite{LipatovEff}, is a non-linear gauge invariant action which correctly reproduces the production of the particles in direct channels at a quasi-multi-Regge kinematics. 
It is written for the local in rapidity interactions of physical gluons in
direct channels inside of some rapidity interval $(y\,-\,\eta/2,y\,+\,\eta/2)$. The interaction between the different clusters of gluons at different though very close rapidities
can be described with the help of reggeized gluon fields
\footnote{We use the Kogut-Soper convention for the light-cone for the light-cone definitions with $x_{\pm}\,=\,\Le x_{0}\,\pm\,x_{3} \Ra/\sqrt{2}$ and $x_{\pm}\,=\,x^{\mp}$\,.} 
$A_{-}$ and $A_{+}$ interacting in crossing channels. Those interaction are non-local in rapidity space. 
This non-local term is not included in the action, the term of interaction between the reggeon fields in the action is local in rapidity and can be considered a 
kind of renormalization term  in the Lagrangian.
The action is gauge invariant and written in the covariant form in terms of gluon field $\textsl{v}$ as
\beq\label{Ef1}
S_{eff}\,=\,-\,\int\,d^{4}\,x\,\Le\,\frac{1}{4}\,F_{\mu \nu}^{a}\,F^{\mu \nu}_{a}\, \,+\,tr\,\left[\,\textsl{v}_{+}\,J^{+}(\textsl{v}_{+})\,-\,A_{+}\,j_{reg}^{+}\,+\,
\,\textsl{v}_{-}\,J^{-}\,(\textsl{v}_{-})\,-\,A_{-}\,j_{reg}^{-}\,\right]\,\Ra\,,
\eeq
where
\beq\label{Ef2}
J^{\pm}(\textsl{v}_{\pm})\,=\,O(x^{\pm}, \textsl{v}_{\pm})\,j_{reg}^{\pm}\,,
\eeq
with $O(x^{\pm}, \textsl{v}_{\pm})$ as some operators, see \cite{LipatovEff},
Appendix A and
\beq\label{Ef3}
j_{reg\,a}^{\pm}\,=\,\frac{1}{C(R)}\,\D_{i}^{2} A_{a}^{\pm}\,,
\eeq
is a reggeon current, where $C(R)$ is the eigenvalue of Casimir operator in the representation R with $C(R)\,=\,N$ in the case of adjoint representation used in the paper.
Further in the calculations we will use the form of the reggeon curnet \eq{Ef3} borrowed from the 
CGC (Color Glass Condensate) approach, see \cite{Venug,Kovner,Hatta1}, where this
current is written in terms of
some color density function defined as
\beq\label{Ef31}
\D_{i}\,\D_{-}\,\rho_{a}^{i}\,=\,-\,\frac{1}{N}\,\D_{\bot}^{2}\,A_{a}^{+}\,,
\eeq
or
\beq\label{Ef32}
\rho_{a}^{i}\,=\,\frac{1}{N}\,\D_{-}^{-1}\,\Le\D^{i}\,A_{-}^{a}\Ra\,,
\eeq
see \cite{Our1} for details.
There are additional kinematical constraints for the reggeon fields
\beq\label{Ef4}
\partial_{-}\,A_{+}\,=\,\partial_{+}\,A_{-}\,=\,0\,,
\eeq
corresponding to the strong-ordering of the Sudakov components in the multi-Regge kinematics,
see \cite{LipatovEff}. Everywhere, as usual,  $\,\partial_{i}\,$ denotes the derivative on transverse coordinates.
Under variation on the gluon fields these currents reproduce the Lipatov's induced currents
\beq\label{Ef5}
\delta\,\Le\,\textsl{v}_{\pm}\, J^{\pm}(\textsl{v}_{\pm})\,\Ra\,=\,\Le \delta\,\textsl{v}_{\pm} \Ra\,j^{ind}_{\mp}(\textsl{v}_{\pm})\,=\,
\Le  \delta\,\textsl{v}_{\pm} \Ra \,j^{\pm}(\textsl{v}_{\pm})\,,
\eeq
with shortness notation $j^{ind}_{\mp}\,=\,j^{\pm}$ introduced.
This current posseses a covariant conservation property:
\beq\label{Ef6}
\Le\,D_{\pm}\,j_{\mp}^{ind}(\textsl{v}_{\pm})\,\Ra^{a}\,=\,\Le D_{\pm}\,j^{\pm}(\textsl{v}_{\pm})\Ra^{a}\,=\,0\,.
\eeq
Here and further we denote the induced current in the component form in the adjoint representation\footnote{We use
$\Le\,T_{a}\,\Ra_{b\,c}\,=\,-\,\imath\,f_{a\,b\,c}$ definition of the matrices and  write only "external" indexes of the $f_{a\,b\,c}\,=\,\Le f_{a}\Ra_{b\,c}$ matrix in the trace. } as 
\beq\label{Ef7}
j_{a}^{\pm}(\textsl{v}_{\pm})\,=\,-\,\imath\,tr[T_{a}\,j^{\pm}(\textsl{v}_{\pm})]\,=\,\frac{1}{N}\,tr\left[\,f_{a}\,O\,f_{b}\,O^{T}\,\right]\,\Le \D_{i}^{2} A_{\mp}^{b} \Ra\,=\,
\frac{1}{N}\,U^{a b}\,\Le\,\D_{i}^{2} A_{\mp}^{b}\,\Ra\,,
\eeq
see \cite{Our1} and Appendixes A as well. It was shown in \cite{Our1}, that if the LO value of classical gluon field in solutions of equations of motion is fixed as
\beq\label{Ef8}
\textsl{v}_{\pm}\,=\,A_{\pm}
\eeq
and if the self-consistency of the solutions is required, as the currents in \eq{Ef1} Lagrangian are reproduced directly in form of \eq{Ef2} without  any additional conditions.
Now, applying  the light-cone gauge $\textsl{v}_{-}\,=\,0$,  the equations of motion can be solved, see the form of the classical solutions in \cite{Our1} and further in the paper.
Therefore, the general expressions for the gluon fields can be written in the following form:
\beq\label{Ef9}
v_{i}^{a}\,\rightarrow\,v^{a}_{i\,cl}\,+\,\ep_{i}^{a}\,,\,\,\,\,v_{+}^{a}\,\rightarrow\,v_{+\,cl}^{a}\,+\,\ep_{+}^{a}\,,
\eeq
where the integration on fluctuations around the classical solutions provides loop corrections to the "net" contribution which is based on the classical solutions only.

\section{Expansion of the Lagrangian around the classical solution}

$\,\,\,\,\,\,$ In this section we consider the first step in construction of the effective action of the approach,
expanding the Lagrangian \eq{Ef1} in terms of the fluctuations and classical fields.  Inserting \eq{Ef9} in the Lagrangian
the only corrections to $g^2$ and $\ep^{2}$ orders will be preserved. This precision provides a one-loop correction to the "net" effective action,
contributions from higher order loops  will be considered in a separate publication.

  The Lagrangian in light-cone gauge has the following form:
\beq\label{TrIn1}
L\,=\,-\,\frac{1}{4}\,F_{ij}^{a}\,F_{ij}^{a}\,+\,F_{i+}^{a}\,F_{i-}^{a}\,+\,\frac{1}{2}\,F_{+-}^{a}\,F_{+-}^{a}\,,
\eeq
where $\,F_{+-}^{a}\,F_{+-}^{a}\,$ term does not consists  transverse fluctuations and $\,F_{ij}^{a}\,F_{ij}^{a}\,$ term does not consist longitudinal fluctuations.

\subsection{The $\,F_{i+}^{a}\,F_{i-}^{a}\,$ term}

$\,\,\,\,\,\,$Inserting expression \eq{Ef9}  in this term 
we obtain: 
\beq\label{1TrIn2}
F_{i+}^{a}\,\rightarrow\,F_{i+}^{a}(v_{+}^{cl},\,v_{i}^{cl})\,+\,\Le D_{i}(v_{i}^{cl})\ep_{+}\Ra^{a}\,-\,\Le D_{+}(v_{+}^{cl})\ep_{i}\Ra^{a}\,+\,
g\,f_{a b c}\,\ep_{i}^{b}\,\ep_{+}^{c}\,,
\eeq
and
\beq\label{1TrIn3}
F_{i-}^{a}\,\rightarrow\,F_{i-}^{a}(v_{i}^{cl})\,-\,\D_{-}\ep_{i}^{a}\,.
\eeq
Therefore we have:
\beqar\label{1TrIn4}
&\,&F_{i+}^{a}\,F_{i-}^{a}\, = \,\Le F_{i+}^{a}\,F_{i-}^{a}\Ra_{cl}\,+\,F_{i-}^{a}(v_{i}^{cl})\,\Le\,\Le D_{i}(v_{i}^{cl})\ep_{+}\Ra^{a}\,-\,\Le D_{+}(v_{+}^{cl})\ep_{i}\Ra^{a}\,\Ra\,-\,
g\,f_{a b c}\,\Le \D_{-} v_{i}^{a\,cl} \Ra\, \ep_{i}^{b}\,\ep_{+}^{c}\,-\,\nonumber \\
&-&\,\Le \D_{-}\ep_{i}^{a} \Ra\,F_{i+}^{a}(v_{+}^{cl},\,v_{i}^{cl})\,-\,\Le \D_{-}\ep_{i}^{a} \Ra\,\Le D_{i}(v_{i}^{cl})\ep_{+}\Ra^{a}\,+\,
\Le \D_{-}\ep_{i}^{a} \Ra\,\Le D_{+}(v_{+}^{cl})\ep_{i}\Ra^{a}\,-\,g\,f_{a b c}\,\Le \D_{-}\ep_{i}^{a} \Ra\,\ep_{i}^{b}\,\ep_{+}^{c}\,.
\eeqar
In this expression we do not account term which is cubic in respect to fluctuations , the linear to fluctuations terms are canceled because of equations of motion. The terms quadratic to transverse fluctuations contribute to the corresponding propagator in the 
Lagrangian and term which is quadratic in respect to combination of transverse and longitudinal fluctuations we write as
\beq\label{1TrIn5}
-\,g\,f_{a b c}\,\Le \D_{-} v_{i}^{a\,cl} \Ra\, \ep_{i}^{b}\,\ep_{+}^{c}\,-\,\Le \D_{-}\ep_{i}^{a} \Ra\,\Le D_{i}(v_{i}^{cl})\ep_{+}\Ra^{a}\,=\,
J_{i}^{a}\,\ep_{i}^{a}\,.
\eeq
Here the current 
\beq\label{1TrIn6}
J_{i}^{a}\,=\,\Le\,\ep_{+}^{a}\,\D_{-}\,\D_{i}\,+\,g\,f_{a b c}\,\ep_{+}^{c}\,\Le \D_{-}\,v_{i}^{b\,cl}\,-\,v_{i}^{b\,cl}\,\D_{-}\Ra\,\Ra\,
\eeq
is some effective current in the Lagrangian.

\subsection{The $\,F_{ij}^{a}\,F_{ij}^{a}\,$ term}

$\,\,\,\,\,\,$We have for this term:
\beq\label{2TrIn1} 
F_{ij}^{a}\,=\,F_{ij}^{a}(v_{i}^{cl})\,+\,\Le D_{i}\ep_{j}\Ra^{a}\,-\,\Le D_{j}\ep_{i}\Ra^{a}\,+\,
g\,f_{abc}\,\ep_{i}^{b}\ep_{j}^{c}\,.
\eeq
Therefore, accounting contributions which are only quadratic to the fluctuations, we obtain:
\beqar\label{2TrIn2}
-\frac{1}{4}\,F_{ij}^{a}\,F_{ij}^{a}\,& = &\,-\frac{1}{4}\,\Le F_{ij}^{a}\,F_{ij}^{a}\Ra_{cl}\,-\,\frac{1}{2}\,
F_{ij}^{a}(v_{i}^{cl})\,\Le \Le D_{i}\ep_{j}\Ra^{a}\,-\,\Le D_{j}\ep_{i}\Ra^{a}\,\Ra\,-\,
\frac{g}{2}\,F_{ij}^{a}(v_{i}^{cl})\,f_{abc}\,\ep_{i}^{b}\ep_{j}^{c}\,-\,\nonumber \\
&-&\,
\frac{1}{2}\,\Le D_{i}\ep_{j}\Ra^{a}\,\Le D_{i}\ep_{j}\Ra^{a}\,+\,
\frac{1}{2}\,\Le D_{i}\ep_{j}\Ra^{a}\,\Le D_{j}\ep_{i}\Ra^{a}\,,
\eeqar
where as usual linear to fluctuations terms do not contribute to the effective action.

\subsection{The $F_{+-}^{a}\,F_{+-}^{a}$ term}

$\,\,\,\,\,\,$This term consists contributions from only longitudinal fluctuations. We have here:
\beq\label{1InIn1}
\frac{1}{2}\,F_{+-}^{a}\,F_{+-}^{a}\,\rightarrow\,\frac{1}{2}\,\Le\,\D_{-}\,v_{+}^{a}\,\Ra\,\Le\D_{-}\,v_{+}^{a}\,\Ra\,=\,-\,\frac{1}{2}\,
v_{+}^{a\,cl}\,\Le\D_{-}^{2}\,v_{+}^{a\,cl}\,\Ra\,-\,\Le\D_{-}^{2}\,v_{+}^{a\,cl}\,\Ra\,\ep_{+}^{a}\,-\,
\frac{1}{2}\,\ep_{+}^{a}\,\Le \D_{-}^{2}\,\ep_{+}^{a}\,\Ra\,.
\eeq
The linear term in \eq{1InIn1} is canceled due the equation of motion, therefore the only first and third terms are considered further.

\subsection{The current term}

$\,\,\,\,\,\,$For the effective current term,
taking into account that the linear to fluctuations term is canceling due the equations of motion, 
we obtain with requested precision the following expansion in terms of longitudinal fluctuations:
\beq\label{4InIn3}
v_{+}^{a}\,J_{a}^{+}(v_{+})\,=\, v_{+}^{a\, cl}\,J_{a\,cl}^{+}(v_{+}^{cl})+\,\frac{1}{2}\,
\Le\frac{\delta^{2}\,\Le\,v_{+}^{a}\,J_{a}^{+}\,\Ra}{\delta v_{+}^{b}\,\delta v_{+}^{c}\,}\Ra_{v_{+}\,=\,v_{+}^{cl}}^{x y}\,
\ep_{+\,x}^{b}\,\ep_{+\,y}^{c}\,.
\eeq
The same current's term we can write as
\beq\label{4InIn31}
v_{+}^{a}\,J_{a}^{+}(v_{+})\,=\,-\,v_{+}^{a\, cl}\,O^{a b}(v_{+}^{cl})\,\Le\D_{i}\,\D_{-}\,\rho_{b}^{i}\Ra\,
-\,\frac{1}{2}\,
\Le\frac{\delta\,U^{b\,a}(v_{+}) }{\delta v_{+}^{c}\,}\Ra_{v_{+}\,=\,v_{+}^{cl}}^{x y}\,\Le\D_{i}\,\D_{-}\,\rho_{a}^{i}\Ra_{x}\,
\ep_{+\,x}^{b}\,\ep_{+\,y}^{c}\,.
\eeq
In order to calculate this expression we have to know the expansion of the following function:
\beq\label{1PIn3}
U^{a\,b}(v_{+})\,=\,tr[\,f_{a}\,O(v_{+})\,f_{b}\,O^{T}(v_{+})]\,
\eeq
with respect to the \eq{Ef9} fluctuation
\beq\label{1PIn5}
v_{+}^{a}\,\rightarrow\,v_{+\,cl}^{a}\,+\,\ep_{+}^{a}\,.
\eeq
Using Appendix A formulas we have:
\beq\label{1PIn6}
U^{a\,b}_{x}(v_{+})\,=\,U^{a\,b}_{x}(v_{+\,0}^{cl})\,+\,g\,\Le U_{1}^{a\, b} \Ra_{x y}^{c}\,
\ep_{+\,y}^{c}\,+\,\frac{1}{2} \,g^{2} \Le U_{2}^{a\, b} \Ra_{x y z}^{c\, d}\,
\ep_{+\,y}^{c}\,\ep_{+\,z}^{d}\,+\,\ldots\,,
\eeq
where the integration on repeating $y,z$ indices is assumed. The coefficients of the expansion read as
\beq\label{1PIn601}
\Le U_{1}^{a\, b} \Ra_{x y}^{c}\,=\,
tr[\,f_{a}\,G^{+}_{x y}\,f_{c}\,O_{y}\,f_{b}\,O^{T}_{x}]\,+\,
tr[\,f_{c}\,G^{+}_{y x}\,f_{a}\,O_{x}\,f_{b}\,O^{T}_{y}]\,
\eeq
and
\beqar\label{1PIn602}
\Le U_{2}^{a\, b} \Ra_{x y z}^{c\, d}\,& = &\,
tr[\,f_{a}\,G^{+}_{x y}\,f_{c}\,G^{+}_{y z}\,f_{d}\,O_{z}\,f_{b}\,O^{T}_{x}]\,+\,
tr[\,f_{a}\,G^{+}_{x z}\,f_{d}\,G^{+}_{z y}\,f_{c}\,O_{y}\,f_{b}\,O^{T}_{x}]\,+\,\nonumber \\
& + &\,
tr[\,f_{d}\,G^{+}_{z x}\,f_{a}\,G^{+}_{x y}\,f_{c}\,O_{y}\,f_{b}\,O^{T}_{z}]\,+\,
tr[\,f_{c}\,G^{+}_{y x}\,f_{a}\,G^{+}_{x z}\,f_{d}\,O_{z}\,f_{b}\,O^{T}_{y}]\,+\,\nonumber \\
& + &
tr[\,f_{d}\,G^{+}_{z y}\,f_{c}\,G^{+}_{y x}\,f_{a}\,O_{x}\,f_{b}\,O^{T}_{z}]\,+\,
tr[\,f_{c}\,G^{+}_{y z}\,f_{d}\,G^{+}_{z x}\,f_{a}\,O_{x}\,f_{b}\,O^{T}_{y}]\,,
\eeqar
see \cite{Our1} and Appendix A for details. 
Therefore we obtain for the \eq{4InIn3} expression:
\beq\label{1PIn7}
v_{+}^{a}\,J_{a}^{+}\,=\,-\,v_{+}^{a\,cl}\,O^{a b}(v_{+}^{cl})\Le\D_{i}\,\D_{-}\,\rho_{b}^{i}\Ra\,-\,
\frac{1}{2}\,g\, \ep_{+\,x}^{a}\,\Le U^{a\,b}_{1} \Ra^{c}_{x\,y} \Le \D_{i} \D_{-} \rho_{b}^{i}\Ra_{x}\,\ep_{+\,y}^{c}\,.
\eeq

\section{One loop effective action: integration over fluctuations}

$\,\,\,\,\,\,$  The computation of the one loop correction to the effective action in light-cone gauge we perform using non-canonical method, integrating out subsequently
transverse and longitudinal fluctuations.	The reason for use of this non-canonical method of calculation is simple. The Lagrangian \eq{Ef1} 
consist new term in comparison to the usual gluon QCD Lagrangian. Consequently, instead canonical equation of motion which relates 
transverse and longitudinal fields, we have the following equation:
\beq\label{OL1}
-\Le D_{i} \Le \partial_{-} \textsl{v}^{i}\,\Ra \Ra_{a}\,-\,\D^{2}_{-}\,\textsl{v}_{a\,+}\,=\,j_{a}^{+}(\textsl{v}_{+})\,
\eeq
see \cite{Our1}. This equation is different from the "canonical" one and will lead to the different constraint in the canonical quantization method, see
\cite{Basseto, Brodsky}. Still, it is possible to make a usual substitution in the Lagrangian which relates these fields, see for example \cite{Brodsky},
and define the canonical light-cone Lagrangian in the usual form in the limit $g\,\rightarrow\,0$\,.
But that will case the shift in the argument of the effective current term, especially in light of condition \eq{Ef8}, and  in turn that will lead to some complicated expression of the 
induced current in the equations of motion to $g^{2}$ and higher orders of perturbative theory.  Therefore, we prefer to use the non-canonical method of introducing of
bare propagators in the theory, calculating the final one-loop expressions in terms of these propagators, see Appendix B. As we see there, after the resummation of one-loop terms,
the well known light-cone propagators, see for example \cite{Kovner,Ayal}, are arising in the expressions. So far it is not clear,
is it result of the chosen precision of the calculations or that is a future of the effective Lagrangian \eq{Ef1}, we will investigate this question in the separate publication.

\subsection{Integration on transverse fluctuations}


$\,\,\,\,\,\,$Collecting quadratic on transverse fluctuations terms and effective current term we obtain:
\beqar\label{3TrIn1}
&-&\frac{1}{2}\ep_{i}^{a}\Le \delta_{ac}\Le \delta_{ij}\,\Box\, +\D_{i}\,\D_{j} \Ra
-2 g f_{abc} \Le \delta_{ij}
\Le v_{k}^{b\,cl} \D_{k}-v_{+}^{b\,cl} \D_{-}\Ra-\frac{1}{2}\Le
v_{j}^{b\,cl} \D_{i} + v_{i}^{b\,cl} \D_{j} - F_{ij}^{b}\Ra  \Ra -
\right.\nonumber\\
&-&\left.\,
\,g^{2}\,f_{a b c_{1}}\,f_{c_{1} b_{1} c}\,
\Le \delta_{i j}\,v_{k}^{b\,cl}\,v_{k}^{b_{1}\,cl}\,
-\,v_{i}^{b\,cl}\,v_{j}^{b_{1}\,cl}\,\Ra\,
\Ra\,\ep_{j}^{c}\,+\,J_{i}^{a}\,\ep_{i}^{a}\,=\,\\
&=&\,-\,\frac{1}{2}\,\ep_{i}^{a}\,\Le\,\Le M_{0} \Ra_{ij}^{ac}\,+\,\Le M_{1} \Ra_{ij}^{ac}\,+\,
\Le M_{2} \Ra_{ij}^{ac}\,\Ra\,\ep_{j}^{c}\,+\,J_{i}^{a}\,\ep_{i}^{a}\,.
\eeqar
There are the following operators with respect to the transverse fluctuations which we determine and which we will use in the further calculations. 
The first one one reads as
\beq\label{3TrIn22}
\tilde{G}_{ij}^{ac}\,=\,\left[\,\Le M_{0} \Ra_{ij}^{ac}\,+\,\Le M_{1} \Ra_{ij}^{ac}\,+\,\Le M_{2} \Ra_{ij}^{ac}\,\right]^{-1}\,,
\eeq
the second one is
\beq\label{3TrIn2}
G_{ij}^{ac}\,=\,\left[\,\Le M_{0} \Ra_{ij}^{ac}\,+\,\Le M_{1} \Ra_{ij}^{ac}\,\right]^{-1}\,,
\eeq
and the the third one, which is the bare propagator of transverse fluctuations,  is
\beq\label{3TrIn3}
G_{0\,ij}^{ac}\,=\,\left[\,\Le M_{0} \Ra_{ij}^{ac}\,\right]^{-1}\,.
\eeq
The inverse operator expressions of \eq{3TrIn22}-\eq{3TrIn22}  we write in the following  perturbative forms:
\beq\label{3TrIn41}
\tilde{G}_{ij}^{ac}(x,y)\,=\,G_{0\,ij}^{ac}(x,y)\,-\,\int\,d^4 z\,G_{0\,ij^{'}}^{ab}(x,z)\,
\Le \,\Le M_{1}(z)\Ra_{j^{'}j^{''}}^{bd}\,+\,\Le M_{2}(z)\Ra_{j^{'}j^{''}}^{bd}\, \Ra \tilde{G}_{j^{''}j}^{dc}(z,y)\,
\eeq
and
\beq\label{3TrIn4}
G_{ij}^{ac}(x,y)\,=\,G_{0\,ij}^{ac}(x,y)\,-\,\int\,d^4 z\,G_{0\,ij^{'}}^{ab}(x,z)\,
\Le M_{1}(z) \Ra_{j^{'}j^{''}}^{bd}\,G_{j^{''}j}^{dc}(z,y)\,,
\eeq
with the bare propagator defined through
\beq\label{3TrIn6}
\Le M_{0}(x) \Ra_{ij}^{ac}\,G_{0\,jl}^{cb}(x\,,y)\,=\,\delta^{ab}\,\delta_{il}\,\delta^{4}(x\,-\,y)\,,
\eeq
where
\beq\label{3TrIn5}
\Le M_{0} \Ra_{ij}^{ac}\,=\,\delta_{ac}\Le \delta_{ij}\,\Box\, +\D_{i}\,\D_{j} \Ra\,.
\eeq
The solution of \eq{3TrIn6}  is simple:
\beq\label{3TrIn7}
G_{0\,ij}^{ab}(x,y)\,=\,-\,\delta^{ab}\,\int\,\frac{d^4 p}{(2\pi)^{4}}\,\frac{e^{-\imath\,p\,(x\,-\,y)}}{p^2}\,\Le
\delta_{ij}\,-\,\frac{p_{i}\,p_{j}}{2\Le p_{-}\,p_{+}\Ra}\Ra\,=\,-\,\delta^{ab}\,G_{0\,ij}(x,y)\,
\eeq
and determines the above operators as perturbative series based on expressions \eq{3TrIn41}-\eq{3TrIn4}.

  Integrating out the transverse fluctuation obtaining the following expression for the effective action: 
\beqar\label{4TrIn1}
\Gamma\,& = &\,\int\,d^{4} x\,\Le L_{YM}(v_{i}^{cl},\, v_{+}^{cl}\,,\ep_{+})- v_{+\,cl}^{a}\,J_{a}^{+}(v_{+}^{cl})- A_{+}^{a}\,\Le\,\D_{i}^{2}\,A_{-}^{a}\,\Ra\,\Ra + \nonumber \\
& + &\,
\frac{\imath}{2}\,\ln\Le\,1\,+\,G_{0}\,M_{1}\,\Ra + 
\frac{\imath}{2}\,\ln\Le\,1\,+\,G\,M_{2}\,\Ra +\nonumber \\
& + &\,
\frac{1}{2}\,\int\,d^{4} x\,\int\,d^{4} y\,\Le\,g\,\ep_{+\,x}^{a}\,\Le U^{a\,b}_{1} \Ra^{c}_{x\,y} \Le \D_{i} \D_{-} \rho_{b}^{i}\Ra_{x}\,\ep_{+\,y}^{c}\,+\,
J_{i\,x}^{a}\,\tilde{G}_{ij}^{ab}(x,y)\,J_{j\,y}^{b}\,\Ra\,,
\eeqar
with all reggeon field terms in the Lagrangian are included.

\subsection{Iintegration on longitudinal fluctuations}

 $\,\,\,\,\,\,$Collecting only quadratic to longitudinal fluctuations terms we write the corresponding part of the action as
\beq\label{1LIn1}
\Gamma_{\ep_{+}^{2}}\,= \,-\frac{1}{2}\,\int\,d^{4} x\,\int\,d^{4} y\,\ep_{+\,x}^{a}\,\Le\,
\Le N_{0}\Ra_{a b}^{+ +}\,+\, \Le N_{1}\Ra_{a b}^{+ +}\,+\,\Le N_{2}\Ra_{a b}^{+ +}\,+\,\Le N_{3}\Ra_{a b}^{+ +}\,
%
\Ra_{x\,y}\,\ep_{+\,y}^{b}\,,
\eeq
with inverse propagator term 
\beq\label{1LIn2}
\Le N_{0}\Ra_{a b}^{+ +}\,=\,\delta_{x y}\,\delta^{a b}\D_{-\,y}^{2}\,-\,\Le\,\D_{-\,x}\,\D_{-\,y}\,\D_{i\,x}\,\D_{j\,y}\,G_{0\,i j}^{a b}(x\,,y)\,\Ra\,.
\eeq
Correspondingly, other terms in the Lagrangian \eq{1LIn1} are determined by the following expressions:
\beq\label{1LIn21}
\Le N_{1}\Ra_{a b}^{+ +}\,=\,-\,g\,\Le U^{a\,c}_{1} \Ra^{b}_{x\,y} \Le \D_{i} \D_{-} \rho_{c}^{i}\Ra_{x}\,,
\eeq
the third term 
\beqar\label{1LIn22}
\Le N_{2}\Ra_{a b}^{+ +} & =&
-\,2\,g\,f_{c d b}\,\Le
\Le \D_{-\,y}\,v_{j\,y}^{d\,cl}\Ra\,\Le \D_{-\,x}\,\D_{i\,x}\,\tilde{G}_{i j}^{a c}(x\,,y)\Ra\,-\,
v_{j\,y}^{d\,cl}\,\Le \D_{-\,y}\, \D_{-\,x}\,\D_{i\,x}\,\tilde{G}_{i j}^{a c}(x\,,y)\Ra\,\Ra\,-\,\nonumber \\
\,& - &\,g^{2}\,f_{c c_1 a}\,f_{d d_{1} b}
\Le\Le \D_{-}\,v_{i}^{c_1\,cl}\Ra\,-\,v_{i}^{c_1\,cl}\,\D_{-}\Ra_{x}\,
\Le\Le \D_{-}\,v_{j}^{d_1\,cl}\Ra\,-\,v_{j}^{d_1\,cl}\,\D_{-}\Ra_{y}\,\tilde{G}_{i j}^{c d}(x,y)\,
\eeqar
and the last one
\beq\label{1LIn23}
\Le N_{3}\Ra_{a b}^{+ +}\,=\, -\,\Le\,\D_{-}\,\D_{i}\,\Ra_{x}\,\Le\,\D_{-}\,\D_{j}\,\Ra_{y}\,\Le\,\tilde{G}_{ij}^{a b}(x,y)\,-\,G_{0\,ij}^{a b}(x,y)\,\Ra\,.
\eeq
The  term \eq{1LIn2} determines the equation for the longitudinal bare propagator:
\beq\label{1LIn3}
\int\,d^{4} y\,N_{0\,a b}^{+ +}(x\,,y)\,G^{b c}_{0\,+ +}(y\,,z)\,=\,\delta^{a c}\,\delta_{x z}\,
\eeq
with solution
\beq\label{1LIn4}
 G^{a b}_{0\,+ +}(x,y)\,=\,-\,2\,
\delta^{ab}\,\int\,\frac{d^4 p}{(2\pi)^{4}}\,\frac{e^{-\imath\,p\,\Le x\,-\,y\Ra}}{p^{2}}\,\frac{p_{+}}{p_{-}}\,=\,-\,\delta^{ab}\,G_{0\,+ +}(x,y)
.
\eeq
Therefore, integrating this fluctuation out, it obtained:
\beqar\label{1LIn5}
\Gamma\,& = &\,\int\,d^{4} x\,\Le L_{YM}(v_{i}^{cl},\, v_{+}^{cl})- v_{+\,cl}^{a}\,J_{a}^{+}(v_{+}^{cl})- A_{+}^{a}\,\Le\,\D_{i}^{2}\,A_{-}^{a}\,\Ra\,\Ra + 
\frac{\imath}{2}\,\ln\Le\,1\,+\,G_{0}\,M_{1}\,\Ra +\nonumber \\
& + &\frac{\imath}{2}\,\ln\Le 1 + G\,M_{2}\Ra +
\frac{\imath}{2}\,\ln\Le 1 + G_{0\,+ +}^{b a}\Le                
\Le N_{1}\Ra_{a b}^{+ +}\,+\Le N_{2}\Ra_{a b}^{+ +}\,+\Le N_{3}\Ra_{a b}^{+ +}\,
\Ra\,\Ra\,,
\eeqar
which is functional of the reggeized gluon fields only.

\section{ RFT calculus based on the effective action}

$\,\,\,\,\,\,$ The construction of the RFT calculus based on effective action \eq{1LIn5}  requires the knowledge of classical solutions of \eq{Ef9} in terms of reggeon fields. This task 
was performed in \cite{Our1}, the found  classical solutions are the following:
\beqar\label{1EAIn1}	
v_{+}^{a\,cl}\,&=&\,A_{+}^{a}\,-\,2\,g\,\Box^{-1}\,\left[\,f_{abc}\,\Le U^{b\,b_1}(A_{+})\,\rho_{b_1}^{i}\Ra\,
\Le \D_{i}\,A_{+}^{c}\Ra\right]\,+\,\nonumber\\
&+&\,
4\,g^2\,\Box^{-1}\,\left[\,f_{abc}\,\Le U^{b\,b_1}(A_{+})\,\rho_{b_1}^{i}\Ra\,
\D_{i}\left\{\Box^{-1}\,\left[
\,f_{c b_{2} c_{1}}\Le U^{b_{2}\,b_{3}}(A_{+})\,\rho_{b_{3}}^{j}\Ra\,\Le \D_{j}\,A_{+}^{c_{1}}\Ra
\right]\right \}
\right]\,=\,\nonumber \\
& = &\,A_{+}^{a}\,+\,g\,\Phi_{+\,1}^{a}(A_{+})\,+\,g^{2}\,\Phi_{+\,2}^{a}(A_{+})\,
\eeqar
and
\beqar\label{2PIn10}
v_{i}^{a\,cl} & = & v_{i 0}^{a}\,+\,g\,v_{i 1}^{a}=
U^{a b}(v_{+}^{cl})\,\rho_{b\, i}\Le x^{-} , x_{\bot}\Ra\,-\,g\,\left[
\Box^{-1}\Le\D^{j}\,P_{j\,i}^{a}+\frac{1}{g}\D_{i}\Le\Le\D^{j} U^{a b}\Ra\rho_{j}^{b}\Ra
+\D_{i}\D_{-}^{-1}\,j^{+}_{a\,1}
\Ra\right]\,=\,\nonumber \\
& = & \,
U^{a b}(v_{+}^{cl})\rho_{b\, i}\,+\,g\,\Lambda_{i\,1}^{a}(A_{+})\,,
\eeqar
with some complicated $P_{j\,i}$ function, see \cite{Our1}, 
and
\beq\label{2PIn31}
j^{+}_{a\,1}\,=\,f_{a b c}\,v_{j\,0}^{b}\,\Le \D_{-}\,v^{j\,c}_{0}\,\Ra\,.
\eeq
On the base of these solutions, the effective action  \eq{1LIn5} can be written as functional of reggeized fields only. We have:
\beq\label{2PIn321}
\Le F_{i+}^{a} F_{i-}^{a}\Ra_{cl} =\,\Le \D_{-}\,v_{i}^{a}\Ra\,\Le D_{+}\,v_{i}\Ra^{a}\,-\,\Le \D_{-}\,v_{i}^{a}\Ra\,\Le \D_{i}\,v_{+}^{a}\Ra\,=\,
v_{+}^{a}\,\Le D_{i}\Le \D_{-}\,v_{i}\Ra\Ra^{a}\,+\,\Le \D_{-}\,v_{i}^{a}\Ra\,\Le \D_{+}\,v_{i}^{a}\Ra\,.
\eeq
Taking into account an identity from equation of motion 
\beq\label{2PIn32}
\Le D_{i}\Le \D_{-}\,v_{i}\Ra\Ra^{a}\,=\,\D_{-}^{2}\,v^{a}_{+}\,+\,U^{a b}(v_{+})\,\Le \D_{-}\,\D_{i}\,\rho_{i}^{b}\,\Ra\,,
\eeq
it is obtained to $g^{2}$ accuracy:
\beq\label{4PIn1}
\Le F_{i+}^{a} F_{i-}^{a}\Ra_{cl} =\,-\, 
\,g^{2}\,\Le\D_{-}\,\Phi_{+\,1}^{a}\Ra\,\Le\D_{-}\,\Phi_{+\,1}^{a}\Ra\,
+\,v_{+}^{a\,cl}\,U^{a b}(A_{+})\,\Le \D_{-}\,\D_{i}\,\rho_{i}^{b}\,\Ra\,
+\,\Le \D_{-}\,v_{i}^{a\,cl}\Ra\,\Le \D_{+}\,v_{i}^{a\,cl}\Ra\,.
\eeq
Correspondingly, there are also the following terms of the "net" effective action: 
\beqar\label{3PIn1}
&\,& \frac{1}{2}\,\Le F_{+-}^{a} F_{+-}^{a}\Ra_{cl} = \,\frac{1}{2} \,g^{2}\,\Le\D_{-}\,\Phi_{+\,1}^{a}\Ra\,
\Le\D_{-}\,\Phi_{+\,1}^{a}\Ra\,=\,
\nonumber \\
&=&\,  2\, g^{2}\,
\Box^{-1}\left[f_{abc}\Le U^{b\,d}(A_{+})\Le\D_{-}\rho_{d}^{i}\Ra\Ra
\Le \D_{i} A_{+}^{c}\Ra\right]\,\Box^{-1}\left[f_{ab_1 c_1}\Le U^{b_1\,d_1}(A_{+})\Le\D_{-}\rho_{d_1}^{j}\Ra\Ra
\Le \D_{j} A_{+}^{c_1}\Ra\right]\,
\eeqar
and
\beq\label{5PIn2}
\,- \,\frac{1}{4}\,\Le F_{i j}^{a} F_{i j}^{a}\,\Ra_{cl}\,=\,- \,\frac{1}{4}\,\Le F_{i j}^{a}\,\Ra_{cl\,0} \Le\,F_{i j}^{a}\,\Ra_{cl\,0}\,
\eeq
with
\beqar\label{5PIn1}
\Le F_{i j}^{a}\Ra_{cl\,0} & = & \rho_{b\, j}\,\D_{i}\, U^{a b}(v_{+\,0}^{cl})\,-\,
\rho_{b\, i}\,\D_{j}\, U^{a b}(v_{+\,0}^{cl})\,+\,
g\,f_{a b c}\,\Le \,U^{b b_1}(v_{+\,0}^{cl})\, \rho_{b_1\, i}\Ra\,
\Le \,U^{c c_1}(v_{+\,0}^{cl})\, \rho_{c_1\, j}\Ra\,+\,\nonumber\\
& + & g\,\Le \D_{i}\,\Lambda_{j\,1}^{a}\,-\,\D_{j}\,\Lambda_{i\,1}^{a}\Ra\,+\,
g^{2}\,f_{a b c}\,\Le
U^{b\, b_{1}}\,\rho_{b_1\, i}\,\Lambda_{j\,1}^{c}\,+\Lambda_{i\,1}^{b}\,U^{c\, c_{1}}\,\rho_{c_1\, j}\,
\Ra\,,
\eeqar
where we notice that the \eq{5PIn2} expression's minimal order is $g^2$.
Now, basing on connection between $A_{-}$ field and $\rho_{i}$ operator
\beq\label{EK101}
\D_{i}\,\D_{-}\,\rho_{a}^{i}\,=\,-\,\frac{1}{N}\,\D_{i}^{2}\,A_{-}\,,
\eeq
or
\beq\label{EK1}
\rho_{i}^{a}(x^{-},\,x_{\bot})\,=\,\frac{1}{N}\,\int\,d^{\,4} z\,G^{-\,0}_{x\,z}\Le\,\D_{i\,z}\,A_{-}^{a}(z^{-},\,z_{\bot})\,\Ra\,,
\eeq
see \cite{Our1} and Appendix A, the effective action \eq{1LIn5} can be expanded in terms of reggeon fields $A_{-}$ and $A_{+}$ as
\beq\label{EK2}
\Gamma\,=\,\sum_{n,m\,=\,0}\,\Le\,A_{+}^{a_{1}}\,\cdots\,A_{+}^{a_{n}}\,K^{a_1\,\cdots\,a_{n}}_{b_1\,\cdots\,b_{m}}\,A_{-}^{b_{1}}\,\cdots\,A_{-}^{b_{m}}\,\Ra\,,
\eeq
that determines this expression as functional of reggeon fields and provides effective vertices of the interactions of the reggeized gluons in the RFT calculus.

\section{Interaction kernels and propagators of reggeized gluons}

$\,\,\,\,\,\,$ Effective action \eq{EK2} can be fully determined in terms of the effective vertices of reggeon fields interactions. Calculating these vertices
one after another  we will reconstruct expression similar to introduced in \cite{Gribov}, see also \cite{Braun, Braun1, Bom, OldRFT, NewRFT1}, which can be considered as QCD Hamiltonian 
for reggeized gluon fields. There are the following well known vertices of inyeractions of reggeon fields: vertex of interaction of $A_{+}$ and $A_{-}$ fields 
and vertex of interaction of two $A_{+}$ fields and two $A_{-}$, which are propagator of reggeized gluons and BFKL kernel correspondingly, see \cite{BFKL};  vertex 
of interaction of six reggeon fields, which can be identified with triple Pomeron vertex, see \cite{TripleV}, or with odderon, see \cite{Odd}. The expression \eq{1LIn5}
consists these vertices plus many other with different precision and different color representations. Calculating the QFT corrections to this effective action we
as well will calculate the corrections to these vertices and will determine the expressions for other complex vertices of interactions of reggeized fields in RFT.
Anyway, a recalculation of the known vertices it is a good test of self-consistency of the effective action. Therefore, in this paper we calculate the
propagator of reggeized gluons, which is the basic element of the small-x BFKL approach, in the framework of the effective action for reggeized gluons.

 The interaction of reggeized gluons $A_{+}$ and $A_{-}$ is defined as effective vertex of interactions of reggeon fields in \eq{EK2}:
\beq\label{EK3}
\Le\,K^{a\,b}_{x\,y}\,\Ra^{+\,-}\,=\,K^{a\,b}_{x\,y}\,=\,\Le\,\frac{\delta^{2}\,\Gamma}{\delta A_{+\,x}^{a}\,\delta A_{-\,y}^{b}}\,\Ra_{A_{+},\,A_{-}\,=\,0}\,,
\eeq
we can call this vertex as interaction kernel as well, see \eq{EK8}-\eq{EK10} below.
The contributions to this kernel are provided by the different terms in the action which are linear with respect to $A_{+},\,A_{-}$ fields.
Namely, the variation of a logarithms in \eq{1LIn5}  gives:
\beqar\label{7PIn01}
-2\,\imath\, K^{a\,b}_{x\,y}\,& = &\,\Le\,\frac{\delta^{2}\,\ln\Le 1 + G\,M\,\Ra}{\delta A_{+\,x}^{a}\,\delta A_{-\,y}^{b}}\,\Ra_{A_{+},\,A_{-}\,=\,0}\,=\,\nonumber \\
&=&\,\left[ \Le \frac{\delta^{2}\,G}{\delta A_{+\,x}^{a}\,\delta A_{-\,y}^{b}}\,M\,+\,\frac{\delta\,G}{\delta A_{+\,x}^{a}}\,
\frac{\delta\,M}{\delta A_{-\,y}^{b}}\,+\,\frac{\delta\,G}{\delta A_{-\,y}^{b}}\,\frac{\delta\,M}{\delta A_{+\,x}^{a}}\,+\,
G\,\frac{\delta^{2}\,M}{\delta A_{+\,x}^{a}\,\delta A_{-\,y}^{b}}\,\Ra\,\Le 1 + G\,M\,\Ra^{-1}\,-\,\right.\nonumber\\
&-&\left.
\Le \frac{\delta\,G}{\delta A_{-\,y}^{b}} M + G\,\frac{\delta\,M}{\delta A_{-\,y}^{b}}\Ra \Le 1 + G\,M\Ra^{-1}
\Le \frac{\delta\,G}{\delta A_{+\,x}^{a}} M + G\,\frac{\delta\,M}{\delta A_{+\,x}^{a}}\Ra \Le 1 + G\,M\Ra^{-1}
\right]_{A_{+},\,A_{-} =\, 0 }\,,
\eeqar
see the expression of the effective action  \eq{1LIn5}.

 Therefore, there are the following contributions in the kernel. The leading order contribution to $ K^{a\,b}_{x y\,0}$ is given by the second term in the \eq{4PIn1} 
and reads as\footnote{We note also, that there are also different kernel, related to $<A_{+}\,A_{+}>$ and $<A_{-}\,A_{-}>$ propagators, in the approach. 
In leading order the contributions to these kernels are zero:
$$
\Le K^{a\,b}_{x y}\Ra_{0}^{+\,+}\,=\,\Le K^{a\,b}_{x\,y}\Ra_{0}^{-\,-}\,=\,0\,,
$$
we do not calculate these propagators in the paper, this task will be considered in separate publication.}
\beq\label{EK4}
 K^{a\,b}_{x y\,0}\,=\,-\,\delta^{a\,b}\,\delta_{x\,y}\,\D_{i\,x}^{2}\,=\,\delta^{a\,b}\,\delta_{x\,y}\,\Le \D_{i}\,\D^{i}\Ra_{x}\,.
\eeq
The NLO and NNLO contributions are determined by the logarithms in the r.h.s. of  \eq{1LIn5}. 
Further the variation of $\rho$ field with respect to $A_{-}$ field will be used as well, we have from \eq{EK1}:
\beq\label{EK5}
\frac{\delta\,\rho_{i}^{a}(x^{-},\,x_{\bot})}{\delta\,A_{-}^{b}(y^{-},\,y_{\bot})}\,=\,
\frac{\delta^{a\,b}}{N}\,\int\,d^{\,4} z\,G^{-\,0}_{x\,z}\,\delta(y^{-}\,-\,z^{-})\,\delta^{2}(y_{\bot}\,-\,z_{\bot})\,\D_{i\,z}\,.
\eeq
Taking into account that
\beq\label{EK6}
G^{-\,0}_{x\,z}\,=\,\tilde{G}^{-\,0}_{x^{-}\,z^{-}}\,\delta(x^{+}\,-\,z^{+})\,\delta^{2}(x_{\bot}\,-\,z_{\bot})\,,
\eeq
we write \eq{EK5} in the following form:
\beq\label{EK61}
\frac{\delta\,\rho_{i}^{a}(x^{-},\,x_{\bot})}{\delta\,A_{-}^{b}(y^{-},\,y_{\bot})}\,=\,
\frac{1}{N}\,\delta^{a\,b}\,\delta^{2}(x_{\bot}\,-\,y_{\bot})\,\tilde{G}^{-\,0}_{x^{-}\,y^{-}}\,\D_{i\,y}\,,
\eeq
where the regularization of pole of $\tilde{G}^{-\,0}_{x^{-}\,y^{-}}$ Green's function depends on it's definition, see Appendix A.

  The propagator corresponding to kernel \eq{EK3} is defined as a solution of the following equation:
\beq\label{EK7}
\int\,d^{4} z\,\Le\,K^{a b}_{x z}\Ra^{-\,+}\,\Le\,D^{b c}_{z y}\Ra_{+\,-}\,=\,\delta^{a c}\,\delta_{x y}\,.
\eeq
For the propagator of the reggeized gluons, $D_{+\,-}$ , the following perturbative series can be defined therefore:
\beq\label{EK8}
\Le D_{x y}^{a c}\Ra_{+\,-}\,=\,\Le D_{x y}^{a c}\Ra_{0\,+\,-}\,-\,\int\,d^{4} z\,\int\,d^{4} w\, \Le D_{x z}^{a b}\Ra_{0\,+\,-}\,
\Le\,\Le\,K^{b d}_{z w}\Ra^{-\,+}\,-\,\Le\,K^{b d}_{z w}\Ra^{-\,+}_{0}\,\Ra\,\Le D_{w y}^{d c}\Ra_{+\,-}\,,
\eeq
or in shorten notations:
\beq\label{EK81}
 D_{x y}^{a c}\,=\, D_{x y\,0}^{a c}\,-\,\int\,d^{4} z\,\int\,d^{4} w\,  D_{x z\,0}^{a b}\,
\Le\,K^{b d}_{z w}\,-\,K^{b d}_{z w\,0}\,\Ra\, D_{w y}^{d c}\,,
\eeq
where 
\beq\label{EK82}
 K^{b d}_{z w}\,=\,\sum_{k\,=\,0}\, K^{b d}_{z w\,k}\,
\eeq
and
\beq\label{EK10}
\int\,d^{4} z\,\,K^{a b}_{x z\,0}\,D^{b c}_{z y\,0}\,=\,\delta^{a c}\,\delta_{x y}\,.
\eeq
The calculation of the NLO kernel $K^{a\,b}_{x\,y\,1}$ is performed in Appendix B and 
 Appendix C. Using \eq{AppC16} and \eq{KPr14}, the following expression for the kernel is obtained:
\beq\label{PlMn3}
-2\,\imath\,K^{a\,b}_{x\,y\,1}\,= \,\frac{\imath\,g^{2}\,N}{4\,\pi}\,\D_{i\,x}^{2}\,\Le\,\int\,\frac{d p_{-}}{p_{-}}\,\int\,\frac{d^{2} p_{\bot}}{(2 \pi)^{2}}\,\int\,\frac{d^{2} k_{\bot}}{(2 \pi)^{2}}\,
\frac{ \,k_{\bot}^{2}}{p_{\bot}^{2}\,\Le\,p_{\bot}\,-\,k_{\bot}\,\Ra^{2}}\,e^{-\imath\,\,k_{i} \,\Le x_{i}\,-\,y_{i}\Ra}\,\Ra\,,
\eeq
where the only leading contributions to the kernel are present, see Appendix C.  
Performing Fourier transform  we write \eq{EK81} as:
\beq\label{PlMn4}
\tilde{D}^{a b}(p_{\bot},\,p_{-})\,=\,\frac{\delta^{a b}}{p_{\bot}^{2}}\,-\,
\frac{g^{2}\,N}{32\,\pi^{3}}\,\int\,\frac{d k_{-}^{'}}{k_{-}^{'}}\,\int\,d^{2} k_{\bot}\,
\frac{ \,p_{\bot}^{2}}{k_{\bot}^{2}\,\Le\,p_{\bot}\,-\,k_{\bot}\,\Ra^{2}}\,\tilde{D}^{a b}(p_{\bot},\,k_{-}^{'})\,,
\eeq
here we used
\beq\label{PlMn5}
D^{a b}_{0}(x,y)\,=\,\delta^{a b}\,\int\,\frac{d^{4} p}{(2\pi)^4}\frac{e^{-\imath\,p\,(x\,-\,y)}}{p_{\bot}^{2}}\,,
\eeq
see \eq{EK4} and \eq{EK10} definitions. Introducing rapidity variable $y\,=\,\frac{1}{2}\,ln\,(\Lambda\,k_{-})$ and taking into account the physical cut-off of the rapidity
related with particles cluster size $\eta$, we obtain after integration on $k_{-}$ variable in the limits $y_{0}\,-\,\eta/2$ and $y_{0}\,+\,\eta/2$:
\beq\label{PlMn6}
\tilde{D}^{a b}(p_{\bot},\eta)\,=\,\frac{\delta^{a b}}{p_{\bot}^{2}}\,-\,
\frac{g^{2}\,N}{16\,\pi^{3}}\,\int^{\eta}_{0}\,d\eta^{'}\,\int\,d^{2} k_{\bot}\,
\frac{ \,p_{\bot}^{2}}{k_{\bot}^{2}\,\Le\,p_{\bot}\,-\,k_{\bot}\,\Ra^{2}}\,\tilde{D}^{a b}(p_{\bot},\eta^{'})\,
\eeq
with
\beq\label{PlMn7}
\epsilon(p_{\bot}^{2})\,=\,-\,\frac{\alpha_{s}\,N}{4\,\pi^{2}}\,\int\,d^{2} k_{\bot}\,
\frac{ \,p_{\bot}^{2}}{k_{\bot}^{2}\,\Le\,p_{\bot}\,-\,k_{\bot}\,\Ra^{2}}\,
\eeq
as intercept of the propagator of reggeized gluons. Rewriting this equation as the differential one:
\beq\label{PlMn8}
\frac{\D\,\tilde{D}^{a b}(p_{\bot},\eta)}{\D\,\eta}\,=\,\tilde{D}^{a b}(p_{\bot},\eta)\,\epsilon(p_{\bot}^{2})\,
\eeq
we obtain the final expression for the propagator:
\beq\label{PlMn9}
\tilde{D}^{a b}(p_{\bot},Y)\,=\,\frac{\delta^{a b}}{p_{\bot}^{2}}\,\Le\, \frac{s}{s_{0}} \,\Ra^{\epsilon(p_{\bot}^{2})}\,
\eeq
with some rapidity interval $0\,<\,\eta\,<\,Y\,=\,\ln(s/s_{0})$ of the problem of interest introduced. We note, that obtained propagator is precisely the well-known one,
see \cite{BFKL}, and thereby we demonstrated the self-consistency of the obtained effective action \eq{1LIn5}. It is interesting to note also, that the 
obtained intercept, the
expression \eq{PlMn7}, is known as well in CGC approach, this is color charge density there, see for example Kovner et al. in \cite{Kovner}.

\section{Conclusion}

$\,\,\,\,\,\,$ The main result of this paper is the expression for one loop effective action for reggeized gluons \eq{1LIn5}. On one hand, this effective action can be considered as the
one loop effective action for gluodynamics with added gauge invariant sourse of longitudinal gluons, calculated on the base of non-trivial classical solutions
for gluon fields. These classical solutions are fully determined in terms of reggeized gluons fields, and, therefore, the same action can be considered as the one loop effective action for reggeized gluons.  The expansion of this action in terms of the reggeons, see expression \eq{EK2}, determines the vertices of interactions of these reggeized fields.  There are all possible vertices
in there, the only limitation is the precision of the expression \eq{1LIn5}. Whereas the $A_{+}$ reggeized field is the argument of the ordered exponential in the classical solutions and
the number of derivative in respect to this field is not limited, see expression \eq{1EAIn1}-\eq{2PIn10} and \cite{Our1}, the $A_{-}$ reggeon field is presented in \eq{1LIn5} in combinations which allows
only limited number of derivations in respect to this field. Therefore, the calculation of the complex vertices related to large number of $A_{-}$ reggeized fields will require
or increase of the precision of the calculations, or calculation of the same effective action in a different gauge, where two types of the ordered exponential, with $A_{+}$
and $A_{-}$ fields in arguments, will be presented in the 
classical solutions. In both cases, the precision of the computations will be determined by the QFT methods, namely it will be limited by the orders of the classical solutions, order of loops included in calculations and by combinations of the fields in the final expression which will survive after the $A_{+},\,A_{-}\,\rightarrow\,0$ limit application. 

 Another important result of the paper is the calculation of the propagator of reggeized gluons \eq{PlMn9} in the framework of the approach. Although this propagator
is well known and widely used in all applications of the effective action, see \cite{EffAct} and \cite{BFKL}, the computation of the propagator fully in the
framework of interest was done first time. This calculation we can consider as the check of the self-consistency of the approach and also as the explanation of the methods
of the calculation of
small-x BFKL based vertices in the framework. There are other important vertices which can be similarly calculated  basing on the expression \eq{1LIn5}.
These verices are  important ingredients of the unitary  corrections to different production and interaction amplitudes of the processes at high energies
and they  will be considered in separate publications.

 As we mentioned above, the expression \eq{1LIn5} describes the interactions of the reggeized gluons inside a cluster of particles in the limited range of rapidity. Therefore,
following to \cite{Gribov}, this expression we can define as RFT Hamiltonian written in terms of the QCD reggeized gluons. The performed calculations, in turn, we can consider as the
construction of 
QCD RFT in terms of QCD degrees of freedom. This RFT construction is interesting because it allows to consider the field theory
in terms of $A_{-}$ and $A_{+}$ fields only, developing approach to the calculation of reggeon loops, applications of the Hamiltonian in the integrable systems frameworks
and in condensed matter physics,
\cite{BFKLInt}, and use of the methods in the effective gravity approach, see \cite{LipatovEff}.

 In conclusion we emphasize, that  the paper is considered as the additional step to the developing of the effective theory for  reggeized gluons which will be 
useful in a variety of applications in high energy physics and other research fields.

\newpage
\section*{Appendix A: Representation and properties of $O$ and $O^{T}$ operators}

\renewcommand{\theequation}{A.\arabic{equation}}
\setcounter{equation}{0}

$\,\,\,\,\,\,$ For the arbitrary representation of gauge field $v_{+}\,=\,\imath\,T^{a}\,v_{+}^{a}$ with
$D_{+}\,=\,\D_{+}\,-\,g\,v_{+}$, we can consider
the following representation of $O$ and $O^{T}$ operators
\footnote{Due the light cone gauge we consider here only $O(x^{+})$ operators. 
The construction of the representation of the $O(x^{-})$ operators can be done similarly. }:
\beq\label{A11}
O_{x}\,=\,\delta^{a\, b}\,+\,g\,\int\,d^{4}y\,G_{x y}^{+\,a\, a_{1}}\, \Le v_{+}(y)\Ra_{a_1\,b} \, =\,
\,1\,+\,g\,G_{x y}^{+}\, v_{+ y}\,
\eeq
and correspondingly
\beq\label{A12}
O^{T}_{x}\,=\,\,1\,+\,g\,v_{+ y}\,G_{y x}^{+}\,,
\eeq
which is redefinition of the operator expansions used in \cite{LipatovEff} in terms of Green's function instead 
integral operators, see \cite{Our1} for more details.
The Green's function in above equations we understand as Green's function of the $D_{+}$ operator
and express it in the perturbative sense as
\beq\label{A13}
G_{x y}^{+}\,=\,G_{x y}^{+\,0}\,+\,g\,G_{x z}^{+\,0}\, v_{+ z}\,G_{z y}^{+}\,
\eeq
and
\beq\label{A14}
G_{y x}^{+}\,=\,G_{y x}^{+\,0}\,+\,g\,G_{y z}^{+}\, v_{+ z}\,G_{z x}^{+\,0}\,,
\eeq
with the bare propagators defined as (there is no integration on $x$ variable)
\beq\label{A15}
\D_{+ x}\,\,G_{x y}^{+\,0}\,=\,\delta_{x\,y}\,,\,\,\,G_{y x}^{+\,0}\,\overleftarrow{\D}_{+ x}\,=\,-\delta_{x\,y}\,.
\eeq
There are the following properties of the operators can be derived:
\begin{enumerate}
\item
\beqar\label{A161}
\delta\,G^{+}_{x y}& = & g\,G_{x z}^{+\,0}\,\Le \delta v_{+ z} \Ra\,G_{z y}^{+}+
\,G_{x z}^{+\,0}\, v_{+ z}\,\delta G_{z y}^{+}=
g\,G_{x z}^{+\,0}\,\Le \delta v_{+ z} \Ra\,G_{z y}^{+}+
\,G_{x z}^{+\,0}\, v_{+ z}\,\Le \delta G_{z p}^{+} \Ra\,D_{+ p}\,G^{+}_{p y}=\nonumber \\
&=&
g \Le
G_{x z}^{+\,0}\,\Le \delta v_{+ z} \Ra\,G_{z y}^{+}-
G_{x z}^{+\,0}\, v_{+ z}\,G_{z p}^{+}\,\Le \delta D_{+ p}\Ra\,G^{+}_{p y}\Ra=
g \Le
G_{x p}^{+\,0}\,+\,G_{x z}^{+\,0}\, v_{+ z}\,G_{z p}^{+}\Ra\,\delta v_{+ p} \,G^{+}_{p y}=\nonumber \\
&=&\,
\,g\,G^{+}_{x p}\,\delta v_{+ p} \,\,G^{+}_{p y}\,.
\eeqar
\item
\beq\label{A16}
\delta\,O_{x}\, = \,g\,G^{+}_{x y}\,\Le \delta v_{+ y} \Ra\,+g\,\Le \delta G^{+}_{x y}\Ra\,v_{+ y}\,=
\,g\,G^{+}_{x p}\,\delta v_{+ p}\,\Le 1\, +\,g \,G^{+}_{p y}\,v_{+ y}\,\Ra\,=\,
g\,G^{+}_{x p}\,\delta v_{+ p}\,O_{p}\,;
\eeq
\item
\beq\label{A17}
\D_{+ x}\,\delta\,O_{x}\,=\,g\,\Le \D_{+ x}\,G^{+}_{x p} \Ra\,\delta v_{+ p}\,O_{p}\,=\,
g\,\Le 1\,+\,g\,\,v_{+ x}\,G^{+}_{x p}\,\Ra\,\delta v_{+ p}\,O_{p}\,=\,
g\,O^{T}_{x}\,\delta v_{+ x}\,O_{x}\,;
\eeq
\item
\beq\label{A18}
\D_{+ x}\,O_{x}\,=\,g\,\Le \D_{+ x}\,G^{+}_{x y} \Ra\,v_{+ y}\,=\,
g\,v_{+ x}\,\Le 1\,+\,g\,G_{x y}^{+}\,v_{+ y}\,\Ra\,=\,
g\,v_{+ x}\,O_{x}\,;
\eeq
\item
\beq\label{A19}
O_{x}^{T}\,\overleftarrow{\D}_{+ x}\,=\,g\,v_{+ y}\,\Le G^{+}_{y x}\,\overleftarrow{\D}_{+ x} \Ra\,=\,
-\,g\,\Le 1\,+\,v_{+ y}\,\,G^{+}_{y x}\,\Ra\,v_{+ x}\,=\,-g\,O^{T}_{x}\,v_{+ x}\,.
\eeq
\end{enumerate}
We see, that the operator $O$ and $O^{T}$ have the properties of ordered exponents. For example, choosing bare propagators as
\beq\label{A110}
\,G_{x y}^{+\,0}\,=\,\theta(x^{+}\,-\,y^{+})\,\delta^{3}_{x y}\,,\,\,\,
\,G_{y x}^{+\,0}\,=\,\theta(y^{+}\,-\,x^{+})\,\delta^{3}_{x y}\,,\,\,\,
\eeq
we immediately reproduce:
\beq\label{A111}
O_{x}\,=\,P\, e^{g\int_{-\infty}^{x^{+}}\,dx^{'+}\, v_{+}(x^{'+})} \,,\,\,\,
O^{T}_{x}\,=\,P\, e^{g\int_{x^{+}}^{\infty}\,dx^{'+}\, v_{+}(x^{'+})} \,.
\eeq
The form of the bare propagators which correspond to another possible integral operator will lead to the
more complicated representations of $O$ and $O^{T}$ operators, see in \cite{LipatovEff}.

  Now we consider a variation of the action's full current :
\beq\label{A112}
\delta\, tr[v_{+ x}\,O_{x}\,\D_{i}^{2}\,A^{+}]=
\frac{1}{g}\,\delta\, tr[\Le \D_{+ x}\,O_{x} \Ra \D_{i}^{2}\,A^{+}]=\frac{1}{g}\, tr[
\Le\D_{+ x} \delta \,O_{x} \Ra \D_{i}^{2}\,A^{+}] = tr[
O^{T}_{x}\,\delta v_{+ x}\,O_{x}\Le \D_{i}^{2}\,A^{+}\Ra]\,,
\eeq 
which can be rewritten in the familiar form of the  Lipatov's induced current used in the paper:
\beq\label{A1121}
\delta\,\Le v_{+}\,J^{+} \Ra\,=\delta\,tr[\, \Le\, v_{+ x}\,O_{x}\,\D_{i}^{2}\,A^{+}\,\Ra\,]\,=\,
\delta v_{+}^{a}\,tr[\,T_{a}\,O\,T_{b}\,O^{T}\,]\,\Le \D_{i}^{2} A^{+}_{b}\Ra\,.
\eeq
We note also, that with the help of \eq{A11} representation of the $O$ operator
the full action's  current can we written as
\beq\label{A113}
 tr[\Le v_{+ x}\,O_{x}\,-\,A_{+}\Ra\,\D_{i}^{2}\,A^{+}\,]\,=\,
tr[\Le v_{+}\,-\,A_{+} + v_{+ x}\,G^{+}_{x y}\,v_{+ y}\,\Ra\,\Le \D_{i}^{2} A^{+}\Ra]\,.
\eeq

\newpage
\section*{Appendix B: Calculation of $ K_{x y\,1}^{a b}$ effective kernel}

\renewcommand{\theequation}{B.\arabic{equation}}
\setcounter{equation}{0}

$\,\,\,\,\,\,$Below we present all contributions to the vertex of interest and present the general answer for the kernel. 
We note, that there are new effective propagators which are arising during the calculations, see \eq{LL15}, \eq{LL21}-\eq{LL22} below. 
Introducing these propagators the effective resummation of the different contributions occurs, see expressions \eq{LL16} and \eq{LL28}. 
This ressumation can be performed directly in the expansion of logarithms in \eq{1LIn5}, before taking the derivatives.
Is that redefinitions of the propagators valid for the all order contributions, i.e is it can be done on the level of the Lagrangian this is a subject of the separate publication, we do not
consider this problem in the paper.

\subsection*{Transverse loop terms contributions}

 $\,\,\,\,\,\,$ First of all, consider the $M_{2}$ term in \eq{1LIn5} to $g^{2}$ accuracy :
\beq\label{7PIn1}
\Le M_{2}\Ra_{i j}^{a b}\,=\,-\,g^2\,
\,f_{a c c_{1}}\,f_{c_{1} c_{2} b}\,
\Le \delta_{i j}\,v_{k\,0}^{c\,cl}\,v_{k\,0}^{c_{2}\,cl}\,
-\,v_{i\,0}^{c\,cl}\,v_{j\,0}^{c_{2}\,cl}\,\Ra\,. 
\eeq
We note that this term is quadratic with respect to $\rho$ field and therefore 
it does not contribute to the kernel of interest.

 For the $M_{1}$ term in \eq{1LIn5} we have:
\beqar\label{7PIn04}
-\,2\,\imath\,\Le K^{a\,b}_{x\,y\,1}\Ra_{1}\, & = &\,\left[ G_{0}\,\frac{\delta^{2}\,\Le M_{1}^{cl}\Ra^{c c }}{\delta A_{+\,x}^{a}\,\delta A_{-\,y}^{b}}\,\Le 1 + G_{0}\,M_{1}\,\Ra^{-1}\,-\,
\right. \nonumber \\
&-&\,\left. G_{0}\,\frac{\delta\,\Le M_{1}^{cl}\Ra^{c d }}{\delta A_{-\,y}^{b}}\Le 1 + G_{0}\,M_{1}\Ra^{-1} G_{0}\,
\frac{\delta\,\Le M_{1}^{cl}\Ra^{d c }}{\delta A_{+\,x}^{a}} \Le 1 + G_{0}\,M_{1}\Ra^{-1}
\right]_{A_{+},A_{-} =\, 0 }\,.
\eeqar
Taking into account that
\beq\label{7PIn51}
M_{1\,i j}^{a b}\,=\,-\,2\,g\,
f_{acb}\,\Le \delta_{ij} \Le v_{k}^{c\,cl}\,\D_{k}-v_{+}^{c\,cl}\,\D_{-}\Ra-
\frac{1}{2}\Le
v_{j}^{c\,cl} \D_{i} + v_{i}^{c\,cl} \D_{j} - \Le F_{ij}^{c}\Ra_{cl}\Ra \Ra\,,
\eeq
with requested accuracy reads as 
\beqar\label{7PIn6}
 M_{1\,i j}^{a b}\,& = &\,-\,2\,g\,f_{a c b}\,\tilde{M}_{1\,i j}^{c\,cl}\,=\,-\,
\,2\,g\,f_{a c b}\,\Le\,U^{c c_1}\,
\delta_{i j}\,\rho_{k}^{c_1}\,\D_{k}\,-\,\frac{1}{2}\,U^{c c_1}\,\Le \rho_{j}^{c_1}\,\D_{i}\,+\,
\rho_{i}^{c_1}\,\D_{j}\,\Ra\,-\,\right.\nonumber\\
&-&\left.
\,\delta_{i j}\,A_{+}^{c}\,\D_{-}\,
\Ra\,\,,
\eeqar
we obtain
\beq\label{7tIn05}
-\,2\,\imath\,\Le K^{a\,b}_{x\,y\,1}\Ra_{1}\, = \,-\,
\int\,d^{4} z\,d^{4} t\,\Le\,
\frac{\delta }{\delta A_{-\,y}^{b}}\, \Le M_{1\,i j}^{c d} \Ra_{z}\,\Ra\,G_{0\,j k}(z,t) \,\Le
\frac{\delta }{\delta A_{+\,x}^{a}} \Le M_{1\,k l}^{d c} \Ra_{t}\Ra\,G_{0\,l i}(t,z)\,.
\eeq

\subsection*{Longitudinal loop terms contribution}

$\,\,\,\,\,\,$There are a few contributions arose from the last logarithm in the expression \eq{1LIn5} . We account only non-zero ones and calculate them one by one.

\subsubsection*{$\Le N_{1}\Ra_{a b}^{+ +}$ term contribution}

 $\,\,\,\,\,\,$The variation of the first term of the last logarithm in  \eq{1LIn5} gives:
\beqar\label{LL1}
&\,&-\,2\,\imath\,\Le K^{a\,b}_{x\,y\,1}\Ra_{2}\, =\,\left[ G_{0\,+ +}\,\frac{\delta^{2}\,\Le N_{1}\Ra_{c c }^{+ +}}{\delta A_{+\,x}^{a}\,\delta A_{-\,y}^{b}}\,\Le 1 + G_{0\,+ +}\,N_{1}^{+ +}\,\Ra^{-1}\,-\,
\right. \nonumber \\
&-&\,\left. G_{0\,+ +}\,\frac{\delta\,\Le N_{1}\Ra_{c d }^{+ +}}{\delta A_{-\,y}^{b}}\Le 1 + G_{0\,+ +}\,N_{1}^{+ +}\Ra^{-1} G_{0\,+ +}\,
\frac{\delta\,\Le N_{1}\Ra_{d c }^{+ +}}{\delta A_{+\,x}^{a}} \Le 1 + G_{0\,+ +}\,N_{1}^{+ +}\Ra^{-1}
\right]_{A_{+},A_{-} =\, 0 }\,,
\eeqar
here we used $G_{0\,+ +}^{a b}\,\rightarrow\,\delta^{a b}\,G_{0\,+ +}$ definition. The $N_{1}$ term, which is determined by \eq{1LIn21}, is quadratic in respect to the reggeon fields, therefore 
the only first term in \eq{LL1}
gives the non-zero contribution:
\beq\label{LL2}
-\,2\,\imath\,\Le K^{a\,b}_{x\,y\,1}\Ra_{2}\, =\,\left[ G_{0\,+ +}\,\frac{\delta^{2}\,\Le N_{1}\Ra_{c c }^{+ +}}{\delta A_{+\,x}^{a}\,\delta A_{-\,y}^{b}}\,\right]_{A_{+},A_{-} =\, 0 }\,.
\eeq
We have:
\beq\label{LL3}
\frac{\delta^{2}\,\Le N_{1}\Ra_{c c }^{+ +}}{\delta A_{+\,x}^{a}\,\delta A_{-\,y}^{b}}\,=\,\frac{g}{N}\,
\frac{\delta\,\Le U_{1}^{c d}\Ra^{c}_{z t}}{\delta A_{+\,x}^{a}}\,
\frac{\delta\,  \D_{i}^{2} A_{-\,z}^{d}}{\delta A_{-\,y}^{b}}\,=\,\frac{g^2}{N}\,
\Le U_{2}^{c b}\Ra^{c a_1}_{z t w}\,\frac{\delta\, v_{+\,w}^{a_1\,cl}}{\delta\,A_{+\,x}^{a}}\,
\Le \delta^{2}_{y_{\bot}\,z_{\bot}} \delta_{y^{-}\,z^{-}}\Ra\,
\D_{i\,z}^{2}\,.
\eeq
At requested accuracy we have:
\beq\label{LL33}
\frac{\delta\, v_{+\,w}^{a_1\,cl}}{\delta\,A_{+\,x}^{a}}\,=\,\delta^{a\,a_1}\,\Le \delta^{2}_{x_{\bot}\,w_{\bot}} \delta_{x^{+}\,w^{+}}\Ra\,,
\eeq
see \eq{1EAIn1} and
\beqar\label{LL4}
\Le\Le U_{2}^{c b}\Ra^{c a}_{z t w}\,\Ra_{A_{+},\,A_{-} = \, 0}\,& = &\,\frac{1}{2}\,N^{2}\,\delta^{a b}\,\left[\,\Le\,
G_{z w}^{+\,0}\,G_{w t}^{+\,0}\,+\,G_{t w}^{+\,0}\,G_{w z}^{+\,0}\,\Ra\,+\,\right.\nonumber \\
&+&\,\left.
\,2\,\Le
G_{z t}^{+\,0}\,G_{t w}^{+\,0}\,+\,G_{z t}^{+\,0}\,G_{w z}^{+\,0}\,+\,G_{z w}^{+\,0}\,G_{t z}^{+\,0}\,+\,G_{t z}^{+\,0}\,G_{w t}^{+\,0}\,
\Ra\,\right]\,.
\eeqar
Therefore, we obtain:
\beqar\label{LL5}
&\,& -\,2\,\imath\,\Le K^{a\,b}_{x\,y\,1}\Ra_{2}\,  = 
 \frac{1}{2}\, g^{2}\,N\,\delta^{a\,b}\,\int d^{4} z\,d^{4} t\, d^{4} w\,\Le\,G_{0\,+ +}^{t z}\, 
\Le \delta^{2}_{x_{\bot}\,w_{\bot}} \delta_{x^{+}\,w^{+}}\Ra\,\Le \delta^{2}_{y_{\bot}\,z_{\bot}} \delta_{y^{-}\,z^{-}}\Ra\,\right.\cdot\nonumber \\
&\cdot& \left. \left[\Le
G_{z w}^{+\,0}\,G_{w t}^{+\,0}+G_{t w}^{+\,0}\,G_{w z}^{+\,0}\Ra+
2\Le
G_{z t}^{+\,0}\,G_{t w}^{+\,0}+G_{z t}^{+\,0}\,G_{w z}^{+\,0}+G_{z w}^{+\,0}\,G_{t z}^{+\,0}+G_{t z}^{+\,0}\,G_{w t}^{+\,0}
\Ra\right] \Ra\D_{i\,z}^{2}\,.
\eeqar

\subsubsection*{$\Le N_{2}\Ra_{a b}^{+ +}$ term contribution}

$\,\,\,\,\,\,$The only contribution from this term  in the vertex of interest comes from the expression
\beq\label{LL61}
-\,2\,\imath\,\Le K^{a\,b}_{x\,y\,1}\Ra_{3}\, =\,\left[ G_{0\,+ +}\,\frac{\delta^{2}\,\Le N_{2}\Ra_{c c }^{+ +}}{\delta A_{+\,x}^{a}\,\delta A_{-\,y}^{b}}\,\right]_{A_{+},A_{-} =\, 0 }\,,
\eeq
which determined by the following derivatives:
\beq\label{LL6}
\frac{\delta^{2}\,\Le N_{2}\Ra_{c c }^{+ +}}{\delta A_{+\,x}^{a}\,\delta A_{-\,y}^{b}}\,=\,
-\,2\,g\,f_{c_1\, d c}\frac{\delta}{\delta\, A_{-\,y}^{b}}\,\Le \Le \D_{-}\,v_{j}^{d\,cl}\Ra\,-\,v_{j}^{d\,cl}\,\D_{-} \Ra_{t}\,
\frac{\delta}{\delta\, A_{+\,x}^{a}}\,\Le \D_{-\,z}\,\D_{i\,z}\,\tilde{G}_{i j}^{c\, c_{1}}(z\,,t)\,\Ra\,.
\eeq
Taking into account that
\beq\label{LL9}
\frac{\delta}{\delta\, A_{+\,x}^{a}}\,\Le \D_{-\,z}\,\D_{i\,z}\,\tilde{G}_{i j}^{c\, c_{1}}(z\,,t)\,\Ra\,=\,-\,
\int\,d^4 w\,\Le \D_{-\,z}\,\D_{i\,z}\,G_{0\,i j_{1}}(z,w) \Ra\,\frac{\delta\,\Le M_{1}(w) \Ra^{c\,c_{1}}_{j_{1}\,j_{2}} }{\delta\, A_{+\,x}^{a}}\,
G_{0\,j_{2} j}(w,t)\,,
\eeq
we collect all terms together obtaining:
\beqar\label{LL10}
-\,2\,\imath\,\Le K^{a\,b}_{x\,y\,1}\Ra_{3}\,& = &\,2\,g\,f_{c_1\, d c}\,\int\,d^{4} t\,d^{4} z\,d^{4} w\,
G_{0\,+ +}(t,z)\,\Le \D_{-\,z}\,\D_{i\,z}\,G_{0\,i j_{1}}(z,w) \Ra\,
\frac{\delta\,\Le M_{1\,j_{1} j_{2}}^{c\,c_{1}} \Ra_{w} }{\delta\, A_{+\,x}^{a}}\,\nonumber \\
&\,&\,
\Le \frac{\delta}{\delta\, A_{-\,y}^{b}}\,\Le \Le \D_{-}\,v_{j}^{d\,cl}\Ra\,-\,v_{j}^{d\,cl}\,\D_{-} \Ra_{t}\,G_{0\,j_{2} j}(w,t)\,\Ra\,.
\eeqar
Here we have from \eq{EK101}
\beq\label{LL7}
\D_{-}\,\rho^{a}_{i}\,=\,\frac{1}{N}\,\D_{i}\,A_{-}^{a}\,,
\eeq
that
\beq\label{LL8}
\frac{\delta}{\delta\, A_{-\,y}^{b}}\,\Le \Le \D_{-}\,v_{j}^{d\,cl}\Ra\,-\,v_{j}^{d\,cl}\,\D_{-}\Ra_{t}\,=\,-\,\delta^{d b}\,\delta^{2}_{y_{\bot}\,t_{\bot}}\,
\Le\,\delta_{y^{-}\,t^{-}}\,-\,\tilde{G}^{-\,0}_{t^{-}\,y^{-}}\,\D_{-\,t}\,\Ra\, \D_{j\,t}\,.
\eeq

\subsubsection*{$\Le N_{3}\Ra_{a b}^{+ +}$ term contribution}

$\,\,\,\,\,\,$The $\Le N_{3}\Ra_{a b}^{+ +}$ term with required accuracy reads as:
\beqar\label{LL11}
 \Le N_{3}\Ra_{a b}^{+ +}\,& = &\,\int\,d^4 z\,\Le \D_{-\,x}\,\D_{i\,x}\,G_{0\,i j_{1}}(x,z) \Ra\,\Le M_{1}(z) \Ra^{a d}_{j_{1}\,j_{2}}\,
\Le \D_{-\,y}\,\D_{j\,y}\,\tilde{G}^{d b}_{j_{2}\,j}(z,y)\Ra\,=\,\nonumber \\
&=&\,
\int\,d^4 z\,\Le \D_{-\,x}\,\D_{i\,x}\,G_{0\,i j_{1}}(x,z) \Ra\,\Le M_{1}(z) \Ra^{a b}_{j_{1}\,j_{2}}\,
\Le \D_{-\,y}\,\D_{j\,y}\,G_{0\,j_{2}\,j}(z,y)\Ra\,-\,\nonumber \\
&-&\,
\int\,d^4 z\,d^{4} t\,\Le \D_{-\,x}\,\D_{i\,x}\,G_{0\,i j_{1}}^{x z}\Ra\,\Le M_{1\,j_{1}\,j_{2}}^{a d} \Ra_{z}\,G_{0\,j_{2}\, j_{3}}^{z t}\,
 \Le M_{1\,j_{3}\,j_{4}}^{d b} \Ra_{t}
\Le \D_{-\,y}\,\D_{j\,y}\,G_{0\,j_{4}\,j}^{t y}\Ra\,, 
\eeqar
here the shorten notations $G_{0\,i j_{1}}(x,z)\,\rightarrow\,G_{0\,i j_{1}}^{x z}$ in the last expression were used and only $ M_{1}$ term was preserved in comparison with \eq{3TrIn41}, the $ M_{2}$ does not 
contribute to the
vertex of interest. Therefore, the first contribution to the kernel is given by the following expression:
\beqar\label{LL12}
&\,&-\,2\,\imath\,\Le K^{a\,b}_{x\,y\,1}\Ra_{4-a}\,= \,\left[ G_{0\,+ +}\,\frac{\delta^{2}\,\Le N_{3}\Ra_{c c }^{+ +}}{\delta A_{+\,x}^{a}\,\delta A_{-\,y}^{b}}\,\right]_{A_{+},\,A_{-} =\, 0 }\,
=\,-\,\left[ G_{0\,+ +}\,\frac{\delta^{2}\,}{\delta A_{+\,x}^{a}\,\delta A_{-\,y}^{b}}\,\nonumber\right. \\
&\,&\left.\Le\,
\int\,d^4 z\,d^{4} t\,\Le \D_{-\,w}\,\D_{i\,w}\,G_{0\,i j_{1}}^{w z}\Ra\,\Le M_{1\,j_{1}\,j_{2}}^{c d} \Ra_{z}\,G_{0\,j_{2}\, j_{3}}^{z t}\,
 \Le M_{1\,j_{3}\,j_{4}}^{d c} \Ra_{t}
\Le \D_{-\,s}\,\D_{j\,s}\,G_{0\,j_{4}\,j}^{t s}\Ra \Ra
\right]_{A_{+},\,A_{-}=0}= \nonumber \\
&=&-2\int\,d^4 w\,d^{4} s\,d^4 z\,d^{4} t\,\Le \,G_{0\,+ +}^{s w}\,
 \Le \D_{-\,w}\,\D_{i\,w}\,G_{0\,i j_{1}}^{w z}\Ra\,\frac{\delta \Le M_{1\,j_{1}\,j_{2}}^{c d} \Ra_{z}}{\delta A_{+\,x}^{a} }\,G_{0\,j_{2}\, j_{3}}^{z t}\,\right. 
\nonumber \\
&\,&\left. \frac{\Le M_{1\,j_{3}\,j_{4}}^{d c} \Ra_{t}}{\delta  A_{-\,y}^{b}}\,
\Le \D_{-\,s}\,\D_{j\,s}\,G_{0\,j_{4}\,j}^{t s}\Ra\,\Ra_{A_{+},\,A_{-} =\, 0 }\,. 
\eeqar
The second contribution of this term to the kernel is given by:
\beq\label{LL13}
-\,2\,\imath\,\Le K^{a\,b}_{x\,y\,1}\Ra_{4-b}\, =\,  -\,\left[\, G_{0\,+ +}\,
\,\frac{\delta\,\Le N_{3}\Ra_{c d }^{+ +}}{\delta A_{-\,y}^{b}}\, G_{0\,+ +}\,
\frac{\delta\,\Le N_{3}\Ra_{d c }^{+ +}}{\delta A_{+\,x}^{a}} \,
\right]_{A_{+},\,A_{-} =\, 0 }\,.
\eeq
Inserting the leading contribution from \eq{LL11} in \eq{LL13}  we obtain:
\beqar\label{LL14}
&\,& -\,2\,\imath\,\Le K^{a\,b}_{x\,y\,1}\Ra_{4-b}\, = \, -\,\int\,d^4 w\,d^{4} s\,d^4 z\,d^{4} t\,\int\,d^{4} w_{1}\,d^{4} t_{1}\,\Le G_{0\,+ +}^{s w}\,
\Le \D_{-\,w}\,\D_{i\,w}\,G_{0\,i j_{1}}^{w\,w_{1}} \Ra\,\Le \frac{\delta}{\delta A_{+\,x}^{a}}\Le M_{1\,j_{1}\,j_{2}}^{c d}\Ra_{w_{1}} \Ra\,\right. \nonumber \\
&\,& \left.\Le \D_{-\,z}\,\D_{j\,z}\,G_{0\,j_{2}\,j}^{w_{1} z}\Ra\,G_{0\,+ +}^{z t}\, \Le \D_{-\,t}\,\D_{l\,t}\,G_{0\,l l_{1}}^{t\,t_{1}} \Ra\,
\Le \frac{\delta}{\delta A_{-\,y}^{b}}\Le M_{1\,l_{1}\,l_{2}}^{d c}\Ra_{t_{1}} \Ra\,
\Le \D_{-\,s}\,\D_{m\,s}\,G_{0\,l_{2}\,m}^{t_{1} s}\Ra\,\Ra_{A_{+},\,A_{-} =\, 0 }\,.  
\eeqar
We introduce now an additional operator: 
\beqar\label{LL15}
\hat{G}_{i j}(x,y)\,& = & \,G_{0\,i\,j}(x,y)\,+\,\int\,d^4 z \,d^4 t\,\Le \D_{-\,z}\,\D_{k\,z}\,G_{0\,i k}(x,z)\Ra\,G_{0\,+ +}(z,t)\,
\Le \D_{-\,t}\,\D_{l\,t}\,G_{0\,l j}(t,y)\Ra\,=\,\nonumber  \\
&=&\, -\,\delta_{i j}\,\int\,\frac{d^4 p}{(2\pi)^{4}}\,\frac{e^{-\imath\,p\,(x-y)}}{p^{2}}\,
\eeqar
and rewrite the following sum
\beq\label{LL16}
\Le K^{a\,b}_{x\,y\,1}\Ra_{1,\,4}\,=\,\Le K^{a\,b}_{x\,y\,1}\Ra_{1}\,+\,\Le K^{a\,b}_{x\,y\,1}\Ra_{4-a}\,+\Le K^{a\,b}_{x\,y\,1}\Ra_{4-b}\,
\eeq
as
\beq\label{LL17}
-\,2\,\imath\,\Le K^{a\,b}_{x\,y\,1}\Ra_{1,\,4}\,=\,-\,\int\,d^{4}z\,d^{4} t\,\,\hat{G}_{i j}^{z t}\,
\Le \frac{\delta}{\delta A_{+\,x}^{a}}\Le M_{1\,j\,j_{1}}^{c d}\Ra_{t} \Ra\,
\hat{G}_{j_{i} j_{2}}^{t z}\,
\Le \frac{\delta}{\delta A_{-\,y}^{b}}\Le M_{1\,j_{2}\,i}^{d c}\Ra_{z} \Ra\,.
\eeq
The functional derivatives in the above expression are given by :
\beq\label{LL171}
\frac{\delta}{\delta A_{+\,x}^{a}}\Le M_{1\,j\,j_{1}}^{c d}\Ra_{t}\,=\,2\,g\,\delta_{j\,j_{1}}\,\delta^{2}_{t_{\bot}\,x_{\bot}}\,\delta_{t^{+}\,x^{+}}\,f_{c a d}\,\D_{-\,t}
\eeq
and
\beq\label{LL172}
\frac{\delta}{\delta A_{-\,y}^{b}}\Le M_{1\,j_{2}\,i}^{d c}\Ra_{z}\,=\,2\,g\,f_{d b c}\, \tilde{G}^{-\,0}_{z^{-}\,y^{-}} \delta^{2}_{z_{\bot}\,y_{\bot}}\,
\Le \delta_{j_{2}\, i}(\D_{\bot\,z})^{2} -\D_{j_{2}\,z} \D_{i\,z}\Ra\,.
\eeq



\subsubsection*{$\Le N_{1}\Ra_{a b}^{+ +}$ and $\Le N_{3}\Ra_{a b}^{+ +}$ terms contribution}

 $\,\,\,\,\,\,$For the non-diagonal contribution from $\Le N_{1}\Ra_{a b}^{+ +}$ and $\Le N_{3}\Ra_{a b}^{+ +}$ terms we have:
\beq\label{LL18}
 -\,2\,\imath\,\Le\,K^{a\,b}_{x\,y\,1}\Ra_{5}\, =\,  -\,\left[\, G_{0\,+ +}\,
\,\frac{\delta\,\Le N_{1}\Ra_{c d }^{+ +}}{\delta A_{-\,y}^{b}}\, G_{0\,+ +}\,
\frac{\delta\,\Le N_{3}\Ra_{d c }^{+ +}}{\delta A_{+\,x}^{a}} \,
\right]_{A_{+},\,A_{-} =\, 0 }\,.
\eeq
Using results from the previous chapters, we obtain:
\beqar\label{LL19}
 -\,2\,\imath\,\Le\,K^{a\,b}_{x\,y\,1}\Ra_{5}\,& = &\,  -\,\frac{g}{N}\,\int\,d^4 w\,d^{4} s\,d^4 z\,d^{4} t\,d^{4} w_{1}\,\Le\,
\,G_{0\,+ +}^{z t}\,\Le\Le U_{1}^{c b}\Ra^{d}_{t w}\Ra_{A_{+},\,A_{-} =\, 0 }\,\Le \delta^{a}_{t_{\bot}\,y_{\bot}}\,\delta_{t^{-}\,y^{-}}\,\D_{i\,t}^{2}\Ra\,\cdot\,\right. \nonumber \\
&\cdot&\,\left.G_{0\,+ +}^{w s}\,
\Le \D_{-\,s}\,\D_{i\,s}\,G_{0\,i j_{1}}^{s w_{1}} \Ra\,\Le\frac{\delta}{\delta A_{+\,x}^{a}}\Le M_{1\,j_{1}\,j_{2}}^{d c}\Ra_{w_{1}}\Ra\,
\Le \D_{-\,z}\,\D_{j\,z}\,G^{w_{1}\,z}_{0\,j_{2}\,j}\Ra\,\Ra\,.
\eeqar
Here we have:
\beq\label{LL20}
\Le\Le U_{1}^{c b}\Ra^{d}_{t w}\Ra_{A_{+},\,A_{-} =\, 0 }\,=\,\frac{1}{2}\,N\,f_{c d b}\,\Le G^{+\,0}_{t w}\,-\,G^{+\,0}_{w t} \Ra\,,
\eeq
see \eq{1PIn601}.
Now we  introduce  new operators:
\beq\label{LL21}
\hat{G}_{+ j}(x,y)\,=\, \int\,d^{4} z\,G_{0\,+ +}(x,z)\,\Le \D_{-\,z}\,\D_{i\,z}\,G_{0\,i j}(z,y) \Ra\,=\,-\,\int\,\frac{d^{4} p}{(2\,\pi)^{4}}
\frac{e^{-\,\imath\,p\,(x\,-\,y)}}{p^{2}}\,\frac{p_{j}}{p_{-}}\,
\eeq
and
\beq\label{LL22}
\hat{G}_{j\, +}(x,y)\,=\,\int\,d^{4} z\,\Le \D_{-\,z}\,\D_{i\,z}\,G_{0\,i j}(x,z) \Ra\, G_{0\,+ +}(z,y)\,\,=\,
\hat{G}_{+ j}(y,x)\,.
\eeq
Finally, with the help of these operators, we rewrite \eq{LL19} as:
\beqar\label{LL23}
 -\,2\,\imath\,\Le\,K^{a\,b}_{x\,y\,1}\Ra_{5}\,& = & \,  -\,\frac{1}{2}\,g\,f_{c d b}\,\int\,d^4 w\,d^{4} t\,d^{4} w_{1}\,\Le\,
\Le G^{+\,0}_{t w}\,-\,G^{+\,0}_{w t} \Ra\,\Le \delta^{a}_{t_{\bot}\,y_{\bot}}\,\delta_{t^{-}\,y^{-}}\,\Ra\,\cdot\,\right.\nonumber \\
& \cdot &\,\left.
\hat{G}_{+ j_{1}}^{w w_{1}}\,\Le\frac{\delta}{\delta A_{+\,x}^{a}}\Le M_{1\,j_{1}\,j_{2}}^{d c}\Ra_{w_{1}}\Ra\,
 \hat{G}_{j_{2}\, +}^{w_{1}\,t}\,\Ra\,\D_{i\,t}^{2}\,.
\eeqar

\subsubsection*{$\Le N_{2}\Ra_{a b}^{+ +}$ and $\Le N_{3}\Ra_{a b}^{+ +}$ terms contribution}

 $\,\,\,\,\,\,$ For the non-diagonal contribution from $\Le N_{2}\Ra_{a b}^{+ +}$ and $\Le N_{3}\Ra_{a b}^{+ +}$ terms we have:
\beq\label{LL24}
 -\,2\,\imath\,\Le\,K^{a\,b}_{x\,y\,1}\Ra_{6}\, =\,  -\,\left[\, G_{0\,+ +}\,
\,\frac{\delta\,\Le N_{2}\Ra_{c d }^{+ +}}{\delta A_{-\,y}^{b}}\, G_{0\,+ +}\,
\frac{\delta\,\Le N_{3}\Ra_{d c }^{+ +}}{\delta A_{+\,x}^{a}} \,
\right]_{A_{+},\,A_{-} =\, 0 }\,.
\eeq
Using results of previous calculations we write:
\beq\label{LL25}
\frac{\delta\,\Le\Le N_{2}\Ra_{c d}^{+ +}\Ra_{t w }}{\delta A_{-\,y}^{b}}\,=\,-\,2\,g\,f_{c_1\, d_{1}\, d}\,
\Le\,\frac{\delta}{\delta\, A_{-\,y}^{b}}\,\Le \Le \D_{-}\,v_{j}^{d_{1}\,cl}\Ra\,-\,v_{j}^{d_{1}\,cl}\,\D_{-} \Ra_{w}\,\Ra\,
\Le \D_{-\,t}\,\D_{i\,t}\,\tilde{G}_{i j}^{c\, c_1}(t\,,w)\,\Ra\,
\eeq
and
\beq\label{LL26}
\frac{\delta\,\Le N_{3}\Ra_{d c }^{+ +}}{\delta A_{+\,x}^{a}} \,=\,\int\,d^{4} w_{1}\,
\Le\D_{-\,s}\,\D_{i\,s}\,G_{0\,i j_{1}}^{s\,w_{1}} \Ra\,\Le \frac{\delta}{\delta A_{+\,x}^{a}}\Le M_{1\,j_{1}\,j_{2}}^{d c}\Ra_{w_{1}} \Ra\,
\Le \D_{-\,z}\,\D_{j\,z}\,G_{0\,j_{2}\,j}^{w_{1} z}\Ra\,.
\eeq
Therefore, we have for the contribution of \eq{LL24} :
\beqar\label{LL27}
&\,&-\,2\,\imath\,\Le\,K^{a\,b}_{x\,y\,1}\Ra_{6}= 2\,g\,f_{c\, d_{1}\, d}\,\int\,d^{4} z\,d^{4} t\,d^{4} w\,d^{4} s\,d^{4} w_{1}\,\Le 
G_{0\,+ +}^{z t} \Le \frac{\delta}{\delta\, A_{-\,y}^{b}} \Le \Le \D_{-}\,v_{j}^{d_{1}\,cl}\Ra - v_{j}^{d_{1}\,cl}\,\D_{-} \Ra_{w}\Ra \right.\nonumber \\
&\,&\left.
\Le \D_{-\,t}\,\D_{i\,t}\,G_{0\,i j}^{t\,w}\,\Ra\, G_{0\,+ +}^{w s}
\Le\D_{-\,s}\,\D_{i\,s}\,G_{0\,i j_{1}}^{s\,w_{1}} \Ra\,\Le \frac{\delta}{\delta A_{+\,x}^{a}}\Le M_{1\,j_{1}\,j_{2}}^{d c}\Ra_{w_{1}} \Ra\,
\Le \D_{-\,z}\,\D_{j\,z}\,G_{0\,j_{2}\,j}^{w_{1} z}\Ra\,\Ra\,=\,\nonumber \\
& = &\,
2\,g\,f_{c\, d_{1}\, d}\,\int\,d^{4} w\,d^{4} w_{1}\,
 \Le \frac{\delta}{\delta\, A_{-\,y}^{b}} \Le \Le \D_{-}\,v_{j}^{d_{1}\,cl}\Ra - v_{j}^{d_{1}\,cl}\,\D_{-} \Ra_{w}\,
\Le \hat{G}^{w_{1} w}_{j_{2} j}\,-\,G_{0\,j_{2} j}^{w_{1} w} \Ra\,\Ra\nonumber \\
&\,&
\hat{G}^{w w_{1}}_{+ j_{1}}
\Le \frac{\delta}{\delta A_{+\,x}^{a}}\Le M_{1\,j_{1}\,j_{2}}^{c d}\Ra_{w_{1}} \Ra\,\,.
\eeqar
Now the following sum
\beq\label{LL28}
\Le\,K^{a\,b}_{x\,y\,1}\Ra_{6,\,3}\,=\,\Le\,K^{a\,b}_{x\,y\,1}\Ra_{6}\,+\,\Le\,K^{a\,b}_{x\,y\,1}\Ra_{3}\,
\eeq
can be written as
\beqar\label{LL281}
-\,2\,\imath\,\Le K^{a\,b}_{x\,y\,1}\Ra_{6,\,3}\, &=&\,
2 g\,f_{c\, d_{1}\, d}\,\int\,d^{4} w d^{4} w_{1}\,
\hat{G}^{w w_{1}}_{+ j_{1}} \Le
 \frac{\delta}{\delta A_{+\,x}^{a}}\Le M_{1\,j_{1}\,j_{2}}^{d c}\Ra_{w_{1}}\Ra\,\cdot \nonumber \\
&\cdot&\,
\Le\,\Le\, \frac{\delta}{\delta\, A_{-\,y}^{b}} \Le \Le \D_{-}\,v_{j}^{d_{1}\,cl}\Ra - v_{j}^{d_{1}\,cl} \D_{-} \Ra_{w}\,\Ra\,
\hat{G}^{w_{1} w}_{j_{2} j}\Ra\,,
\eeqar
with the derivatives in the expression given by \eq{LL8} and \eq{LL171}.

\subsection*{Final expression for the kernel}

 $\,\,\,\,\,\,$Taking all contributions together we obtain the following expression for the kernel to required order:
\beqar\label{LL29}
 &\,&-\,2\,\imath\, K^{a\,b}_{x\,y\,1}\,= \,
\frac{1}{2}\, g^{2}\,N\,\delta^{a\,b}\,\int d^{4} z\,d^{4} t\, d^{4} w\,\D_{i\,z}^{2}\,\Le G_{0\,+ +}^{t z}\, 
\Le \delta^{2}_{x_{\bot}\,w_{\bot}} \delta_{x^{+}\,w^{+}}\Ra\,\Le \delta^{2}_{y_{\bot}\,z_{\bot}} \delta_{y^{-}\,z^{-}}\Ra\,\right.\cdot\nonumber \\
&\cdot&\left. \left[\,\Le\,
G_{z w}^{+\,0}\,G_{w t}^{+\,0}\,+\,G_{t w}^{+\,0}\,G_{w z}^{+\,0}\,\Ra\,+
\,2\,\Le
G_{z t}^{+\,0}\,G_{t w}^{+\,0}\,+\,G_{z t}^{+\,0}\,G_{w z}^{+\,0}\,+\,G_{z w}^{+\,0}\,G_{t z}^{+\,0}\,+\,G_{t z}^{+\,0}\,G_{w t}^{+\,0}\,
\Ra\,\right]\,\Ra\,-\,\nonumber \\
&-&\,
\,\int\,d^{4}z\,d^{4} t\,\,\hat{G}_{i j}^{z t}\,
\Le \frac{\delta}{\delta A_{+\,x}^{a}}\Le M_{1\,j\,j_{1}}^{c d}\Ra_{t} \Ra\,
\hat{G}_{j_{i} j_{2}}^{t z}\,
\Le \frac{\delta}{\delta A_{-\,y}^{b}}\Le M_{1\,j_{2}\,i}^{d c}\Ra_{z} \Ra\,-\,\nonumber \\
&-&\,
\frac{1}{2}\,g\,f_{c d b}\,\int\,d^4 w\,d^{4} t\,d^{4} w_{1}\,\D_{i\,t}^{2}\,\Le\,
\Le G^{+\,0}_{t w}\,-\,G^{+\,0}_{w t} \Ra\,\Le \delta^{2}_{t_{\bot}\,y_{\bot}}\,\delta_{t^{-}\,y^{-}}\,\Ra\,
\hat{G}_{+ j_{1}}^{w w_{1}}\,\Le\frac{\delta}{\delta A_{+\,x}^{a}}\Le M_{1\,j_{1}\,j_{2}}^{d c}\Ra_{w_{1}}\Ra\,
\hat{G}_{j_{2}\, +}^{w_{1}\,t}\Ra +\,\nonumber \\
&+&\,
2 g f_{c\, d_{1}\, d}\int d^{4} w  d^{4} w_{1}
\hat{G}^{w w_{1}}_{+ j_{1}}\Le
 \frac{\delta}{\delta A_{+\,x}^{a}}\Le M_{1\,j_{1}\,j_{2}}^{d c}\Ra_{w_{1}}\Ra
\Le\frac{\delta}{\delta  A_{-\,y}^{b}} \Le \Le \D_{-}\,v_{j}^{d_{1}\,cl}\Ra - v_{j}^{d_{1}\,cl} \D_{-} \Ra_{w}\,\Ra
\hat{G}^{w_{1} w}_{j_{2} j} ,
\eeqar
where the expressions for the all functional derivatives are determined above. 

\newpage
\section*{Appendix C: Calculation of the final answer for $ K_{x y\,1}^{a b}$ effective kernel}

\renewcommand{\theequation}{C.\arabic{equation}}
\setcounter{equation}{0}

$\,\,\,\,\,\,$Below we present final answer for the vertex of interest after the calculations of integrals in \eq{LL29}. We note that
the second and fourth integrals in \eq{LL29} are a contribution of the usual gluon field (to one-loop precision) to the propagator 
of the reggeized gluons, whereas the first and third terms represent new contributions to the propagator leading at high-energy limit.

 Calculating  integrals of \eq{LL29}, we remind, that in the framework of approach
we consider the cluster of the particles, where we can limit the integration on $p_{-}$ variable by $p_{-}\,>\,0$ region. It corresponds to the integration over
$0\,<\,y\,<\,\eta$ limits in the integral over rapidity. In this case we have:
\beq\label{KPr7}
\frac{1}{p^{2}}\,\rightarrow\,\frac{1}{2\,p_{-}\,\Le p_{+}\,-\,\frac{p_{\bot}^{2}}{2\,p_{-}}\,+\,\imath\,\varepsilon\Ra} \,,\,\,\,\,\varepsilon\,>\,0\,,
\eeq
that determines the form of the integration contours in the $p_{+}$ integrals.
Correspondingly, in this kinematic region the usual QCD one-loop gluon contribution, i.e. 
the second and fourth integrals in \eq{LL29}, are zero.
In general, an expansion of the value of $p_{-}$ variable to the  $p_{-}\,<\,0$ region in integrals of \eq{LL29}, will lead to the change 
to the limits of integration in integrals over rapidity only:
\beq\label{KPr71}
\int_{0}^{\eta}\,d\eta^{'}\,\rightarrow\,\int_{0}^{\eta}\,d\eta^{'}\,+\,\int^{0}_{-\eta}\,d\eta^{'}\,=\,\int_{-\eta}^{\eta}\,d\eta^{'}\,,
\eeq
and will not change the integrals over transverse momenta, these integrals are symmetrical with respect to sign of
$p_{-}$ and the final expressions there do not depend on it. In this case the usual gluon one-loop contribution will be not zero anymore, but
it will be sub-leading in the high energy regime, therefore we will not consider it here.

\subsection*{Non-zero contribution: first term of \eq{LL29}}

  For the $p_{-}\,>\,0$ integration limit, the non-vanishing contributions in this term read as:
\beqar\label{AppC11}
I_{1} & = &\frac{1}{2}\, g^{2}\,N\,\delta^{a\,b}\,\int d^{4} z\,d^{4} t\, d^{4} w\,\Le \D_{i\,z}^{2}\,G_{0\,+ +}^{t z} \Ra\, 
\Le \delta^{2}_{x_{\bot}\,w_{\bot}} \delta_{x^{+}\,w^{+}}\Ra\,\Le \delta^{2}_{y_{\bot}\,z_{\bot}} \delta_{y^{-}\,z^{-}}\Ra\,
\Le \,G_{t w}^{+\,0}\,G_{w z}^{+\,0}\,\right. +\, \\
&+&\left.\,2\,\Le \,G_{z w}^{+\,0}\,G_{t z}^{+\,0}\,+\,G_{t z}^{+\,0}\,G_{w t}^{+\,0}\, \Ra\,\Ra\,
\eeqar
which can be written as following:
\beqar\label{AppC12}
I_{1} & = & g^{2}\,N\,\delta^{a\,b}\,\D_{i\,x}^{2} \Le \,\delta^{2}_{x_{\bot}\,y_{\bot}}\, \int   d t^{+}\,\theta(t^{+}\,-\,x^{+})\int\,d z^{+}\,\theta(x^{+}\,-\,z^{+}) \,\right.\nonumber\\
&\,&\left.\int\,d p_{-} \int d p_{+} \int\,\frac{d^{2} p_{\bot}}{(2 \pi)^{4}} \frac{p_{+}}{p^{2}\,p_{-}}\,e^{-\,\imath\,p_{+}(t^{+}\,-\,z^{+})}\Ra\,,
\eeqar
see results of Appendix A, \eq{1LIn4} and  Kovner et al. in \cite{Kovner}.
After the regularization of the theta functions we have:
\beq\label{AppC13}
\int   d t^{+}\,\theta(t^{+}\,-\,x^{+})\,e^{-\,\imath\,p_{+}\,t^{+}}\,=\,-\,\frac{\imath}{p_{+}\,-\,\imath\,\varepsilon}\,e^{-\,\imath\,p_{+}x^{+}}\,
\eeq
and
\beq\label{AppC14}
\int   d z^{+}\,\theta(x^{+}\,-\,z^{+})\,e^{\,\imath\,p_{+}\,z^{+}}\,=\,-\,\frac{\imath}{p_{+}\,-\,\imath\,\varepsilon}\,e^{\imath\,p_{+}x^{+}}\,.
\eeq
Inserting these expression  in \eq{AppC12} we obtain:
\beq\label{AppC15}
I_{1}=-\,g^{2}\,N\,\delta^{a\,b}\,\D_{i\,x}^{2} \Le \, \delta^{2}_{x_{\bot}\,y_{\bot}}\,
\int\,\frac{d p_{-}}{2\,p_{-}^{2}}\,\int\,\frac{d^{2} p_{\bot}}{(2 \pi)^{4}}\,\int d p_{+}\,
\frac{1}{p_{+}\,-\,\frac{p_{\bot}^{2}}{2\,p_{-}}\,+\,\imath\varepsilon}\,\frac{1}{p_{+}\,-\,\imath\varepsilon}\,\Ra\,,
\eeq
that after the integration on $p_{+}$ gives finally:
\beq\label{AppC16}
I_{1}=\,\imath\,\frac{g^{2}\,N}{(2\pi)^{3}}\,\delta^{a\,b}\,\D_{i\,x}^{2} \Le \,
\int\,\frac{d p_{-}}{p_{-}}\,\int\,\frac{d^{2} p_{\bot}}{p_{\bot}^{2}}\,\int\,\frac{d^{2} k_{\bot}}{(2\pi)^{2}}\,e^{-\imath\,k_{\bot}\,\Le x_{\bot}\,-\,y_{\bot} \Ra}\Ra\,.
\eeq

\subsection*{Non-zero contribution: third term of \eq{LL29}}

$\,\,\,\,\,\,$In the $p_{-}\,>\,0$ integration limit, the non-vanishing contribution reads as:
\beq\label{KPr1}
I_{2}=
\frac{1}{2}\,g\,f_{c d b}\,\int\,d^4 w\,d^{4} t\,d^{4} w_{1}\,
G^{+\,0}_{w t}\Le \delta^{2}_{t_{\bot}\,y_{\bot}}\,\delta_{t^{-}\,y^{-}}\Ra
\hat{G}_{+ j_{1}}^{w w_{1}}\,\Le\frac{\delta}{\delta A_{+\,x}^{a}}\Le M_{1\,j_{1}\,j_{2}}^{d c}\Ra_{w_{1}}\Ra
\Le \D_{i\,t}^{2}\,\hat{G}_{j_{2}\, +}^{w_{1}\,t}\Ra\,.
\eeq
Here we have:
\beq\label{KPr2}
\frac{\delta}{\delta A_{+\,x}^{a}}\Le M_{1\,j_{1}\,j_{2}}^{d c}\Ra_{w_{1}}\,=\,2\,g\,\delta_{j_{1}\,j_{2}}\,\delta^{2}_{w_{1\,\bot}\,x_{\bot}}\,\delta_{w_{1}^{+}\,x^{+}}\,f_{d a c}\,\D_{-\,w_{1}}\,,
\eeq
and
\beq\label{KPr3}
G^{+\,0}_{w t}\,=\,\theta(w^{+}\,-\,t^{+})\,\delta^{2}_{w_{\bot}\,t_{\bot}}\,\delta_{w^{-}\,t^{-}}\,,
\eeq
see \eq{LL171} and Appendix A.
Inserting these expressions in \eq{KPr1} we obtain for the first contribution:
\beqar\label{KPr4}
I_{2}\,& = &\,g^{2}\,N\,\delta^{a\,b}\,\int\,d w^{+}\,\theta(w^{+}\,-\,t^{+})\,\int\,dt^{+}\,\int\,d w_{1}^{+}\,\delta_{w_{1}^{+}\,x^{+}}
\int\,d w^{-}\,\delta_{w^{-}\,t^{-}}\,\int\,d t^{-}\,\delta_{t^{-}\,y^{-}}\,\\
&\,&\,\int\,d w^{-}_{1}\,\int\,d^{2} w_{\bot}\,\delta^{2}_{w_{\bot}\,t_{\bot}}\,\int\,d^{2} t_{\bot}\,\delta^{2}_{t_{\bot}\,y_{\bot}}\,
\int\,d^{2} w_{1\,\bot}\,\delta^{2}_{w_{1\,\bot}\,x_{\bot}}\,\hat{G}_{+ j}^{w w_{1}}\,\Le \D_{-\,w_{1}}\,\D_{i\,t}^{2}\,\hat{G}_{j\, +}^{w_{1}\,t}\Ra\,.
\eeqar
The propagator in \eq{KPr4} is defined above, see \eq{LL21}, it is:
\beq\label{KPr5}
\hat{G}_{+ j}^{w w_{1}}\,=\,-\,\int\,\frac{d^{4} p}{(2\pi)^{4}}\,\frac{e^{-\imath\,p\,(w\,-\,w_{1})}}{p^{2}}\,\frac{p_{j}}{p_{-}}\,
\eeq
and correspondingly
\beq\label{KPr6}
\D_{-\,w_{1}}\,\D_{i\,t}^{2}\,\hat{G}_{j\, +}^{w_{1}\,t}\,=\,-\,\imath\,\D_{i\,t}^{2}\,\int\,\frac{d^{4} k}{(2\pi)^{4}}\,e^{\imath\,k\,(w_{1}\,-\,t)}\,
\frac{k_{j}}{k^2}\,.
\eeq

Now, performing integration on delta-functions, we obtain:
\beqar\label{KPr8}
I_{2}\,& = &\,\imath\,g^{2}\,N\,\delta^{a\,b}\,\D_{i\,x}^{2}\,\Le\,\int\,dt^{+}\,\int\,d w^{+}\,\theta(w^{+}\,-\,t^{+})\,\int\,d w^{-}_{1}\,e^{\imath\,w_{1}^{-}\,\Le p_{-}\,+\,k_{-} \Ra}\,
\int\,\frac{d^{4} p}{(2\pi)^{4}}\,\right.\,\\
&\,&\,\left.\,\int\,\frac{d^{4} k}{(2\pi)^{4}}\,\frac{p_{j}}{p_{-}\,p^{2}}\,\frac{k_{j}}{k^2}
e^{-\imath\,y^{-}\,\Le p_{-}\,+\,k_{-} \Ra}\,e^{-\imath\,\Le x_{i}\,-\,y_{i} \Ra\,\Le p_{i}\,+\,k_{i} \Ra}\,e^{-\imath\,p_{+}\,\Le w^{+}\,-\,x^{+} \Ra}
e^{\imath\,k_{+}\,\Le x^{+}\,-\,t^{+} \Ra}\Ra .
\eeqar
The integral on $w_{1}^{-}$ variable provides
\beq\label{KPr9}
\int\,d w^{-}_{1}\,e^{\imath\,w_{1}^{-}\,\Le p_{-}\,+\,k_{-} \Ra}\,=\,2\,\pi\,\delta_{p_{-}\,\,-k_{-}}\,,
\eeq
and integration on $w^{+}$ variable gives
\beq\label{KPr10}
\int\,d w^{+}\,\theta(w^{+}\,-\,t^{+})\,e^{-\imath\,p_{+}\,w^{+}}=\,\int_{t^{+}}^{\infty}\,\,d w^{+}\,e^{-\imath\,w^{+}\,\Le\,p_{+}\,-\,\imath\,\varepsilon\Ra\,}\,=\,-\,
\frac{\imath}{p_{+}\,-\,\imath\,\varepsilon}\,e^{-\imath\,p_{+}\,t^{+}}\,.
\eeq
Therefore, we obtain for \eq{KPr8}:
\beqar\label{KPr11}
I_{2}\,& = &\,-\,2\pi\,g^{2}\,N\,\delta^{a\,b}\,\D_{i\,x}^{2}\,\Le\,
\int\,d k_{+}\,\int\,dt^{+}\,e^{-\imath\,t^{+}\,\Le p_{+}\,+\,k_{+} \Ra}\,\int\,\frac{d p_{-}}{p_{-}}\,\int\,d p_{+}\,
\int\,\frac{d^{2} p_{\bot}}{(2 \pi)^{4}}\,\int\,\frac{d^{2} k_{\bot}}{(2 \pi)^{4}}\,\frac{1}{p_{+}\,-\,\imath\,\varepsilon}\,\nonumber \right.\\
&\,&\,\left.\frac{p_{j}}{2\,p_{-\,}\Le p_{+}\,-\,\frac{p_{\bot}^{2}}{2\,p_{-}}\,+\,\imath\,\varepsilon\Ra}\,
\frac{k_{j}}{2\,p_{-\,}\Le k_{+}\,+\,\frac{k_{\bot}^{2}}{2\,p_{-}}\,-\,\imath\,\varepsilon\Ra}\,
e^{-\imath\,\Le p_{i}\,-\,k_{i} \Ra\,\Le x_{i}\,-\,y_{i}\Ra}\,e^{\imath\,x^{+}\,\Le k_{+}\,+\,p_{+} \Ra}\,\Ra\,.
\eeqar
Integrating now on $t^{+}$ and $k_{+}$ variables we have
\beqar\label{KPr12}
I_{2}\,& = &\,(2\pi)^{2}\,g^{2}\,N\,\delta^{a\,b}\,\D_{i\,x}^{2}\,\Le\,\int\,\frac{d p_{-}}{p_{-}}\,\int\,\frac{d^{2} p_{\bot}}{(2 \pi)^{4}}\,\int\,\frac{d^{2} k_{\bot}}{(2 \pi)^{4}}\,
\Le p_{j}\,k_{j} \Ra\,e^{-\imath\,\Le p_{i}\,+\,k_{i} \Ra\,\Le x_{i}\,-\,y_{i}\Ra}\,\right.\nonumber\\
&\,&\,\left.\int\,\frac{d p_{+}}{4 \,p_{-}^{2}}\,\frac{1}{\Le\, p_{+}\,-\,\frac{k_{\bot}^{2}}{2\,p_{-}}\,+\,\imath\,\varepsilon\,\Ra}\,
\frac{1}{\Le\, p_{+}\,-\,\frac{p_{\bot}^{2}}{2\,p_{-}}\,+\,\imath\,\varepsilon\,\Ra}\,\frac{1}{\Le\,p_{+}\,-\,\imath\,\varepsilon\,\Ra}\,\Ra\,,
\eeqar
and  performing $p_{+}$ integration we obtain:
\beq\label{KPr13}
I_{2}\,= \,\frac{\imath\,g^{2}\,N}{2\,\pi}\,\delta^{a\,b}\,
\D_{i\,x}^{2}\,\Le\,\int\,\frac{d p_{-}}{p_{-}}\,\int\,\frac{d^{2} p_{\bot}}{(2 \pi)^{2}}\,\int\,\frac{d^{2} k_{\bot}}{(2 \pi)^{2}}\,
\frac{ p_{j}\,k_{j}}{p_{\bot}^{2}\,k_{\bot}^{2}}\,e^{-\imath\,\Le p_{i}\,+\,k_{i} \Ra\,\Le x_{i}\,-\,y_{i}\Ra}\,\Ra\,.
\eeq
Performing the variable change $p_{\bot}\,+\,k_{\bot}\,\rightarrow\,k_{\bot}$ we rewrite the integral in the more familiar form:
\beqar\label{KPr14}
I_{2}\,& = & \,\imath\,\delta^{a\,b}\,\frac{g^{2}\,N}{4\,\pi}\,\D_{i\,x}^{2}\,\Le\,\int\,\frac{d p_{-}}{p_{-}}\,\int\,\frac{d^{2} p_{\bot}}{(2 \pi)^{2}}\,\int\,\frac{d^{2} k_{\bot}}{(2 \pi)^{2}}\,
\frac{ \,k_{\bot}^{2}}{p_{\bot}^{2}\,\Le\,p_{\bot}\,-\,k_{\bot}\,\Ra^{2}}\,e^{-\imath\,\,k_{i} \,\Le x_{i}\,-\,y_{i}\Ra}\,\Ra\,-\,\nonumber \\
& - &\,
\imath\,\delta^{a\,b}\,\frac{g^{2}\,N}{(2\,\pi)^{3}}\,\D_{i\,x}^{2}\,\Le\,\int\,\frac{d p_{-}}{p_{-}}\,\int\,\frac{d^{2} p_{\bot}}{p_{\bot}^{2}}\,\int\,\frac{d^{2} k_{\bot}}{(2\,\pi)^{2}} \,
\,e^{-\imath\,\,k_{i} \,\Le x_{i}\,-\,y_{i}\Ra}\,\Ra\,.
\eeqar

\subsection*{Other contributions to \eq{LL29}}

 Other terms in \eq{LL29} will give non-zero contributions in the kernel in the $p_{-}\,<\,0$ integration region only, see discussion above.

\end{document}